\documentclass[twocolumn]{autart}    
\usepackage{float}
\usepackage{mathrsfs}
\usepackage{amsfonts} 
\usepackage{subfigure}
\usepackage{latexsym}
\usepackage{amssymb}
\usepackage{amsfonts}
\usepackage{amsmath}
\usepackage{amsfonts}
\usepackage{graphicx}
\usepackage{color}
\usepackage{mathrsfs}
\usepackage{graphicx}          

\begin{document}

\begin{frontmatter}

\title{Fast state estimation under sensor attacks: a senor categorization approach\thanksref{footnoteinfo}} 

\thanks[footnoteinfo]{This paper was not presented at any IFAC
meeting. Corresponding author: Guang-Hong~Yang. Tel. +XXXIX-VI-mmmxxi.
Fax +XXXIX-VI-mmmxxv.}

\author[Paestum]{Liwei An}\ead{liwei.an@foxmail.com},    
\author[Paestum,Rome]{Guang-Hong Yang}\ead{yangguanghong@ise.neu.edu.cn}               

\address[Paestum]{College of Information Science and Engineering, Northeastern University, Shenyang 110819, P.R.China}  
\address[Rome]{State Key Laboratory of Synthetical Automation
for Process Industries, Northeastern University, Shenyang 110819, P.R.China}


\begin{abstract}                          
In a sensor network, some sensors usually provide the same or equivalent measurement information, which is not taken into account by the existing secure state estimation methods against sparse sensor attacks such that the computational efficiency of these methods needs to be further improved. In this paper, by considering the observation equivalence of sensor measurement, a concept of {\it analytic sensor types} is introduced based on the equivalent class to develop a fast state estimation algorithm. By testing the measurement data of a sensor type, the attack location information can be extracted to exclude some mismatching search candidates, without loss of estimation correctness. This confirms high speed performance of the proposed algorithm, since the number of sensor types is usually far less than the number of sensors.
\end{abstract}

\end{frontmatter}

\section{Introduction}


Recently, the secure state estimation (SSE) problem, aiming at reconstructing the true system from the arbitrarily corrupted measurements, has received a wide attention in the security issue of cyber-physical systems (CPSs) \cite{SS2012}. Although the attack signal can usually be assumed to be sparse due to the execution resource limit of an attacker itself, the large attack space still makes these SSE algorithms inevitably complicated. Due to intrinsically combinatorial complexity, the study on SSE problem becomes more important on computation and memory limited platforms. Exhaustive search as a simple state estimation algorithm has been well studied \cite{CL2015,FP2013,MS2015,SM2014}. However, this approach is typically computationally intractable and cannot be implemented in practice since the combinatorial problem has to be solved repeatedly.

In the literature, many representative results on SSE have appeared for finding computationally efficient algorithms. In \cite{YH2015,CS2017,HF2014,MP2017}, an $L_1/L_r$ relaxation technique is devoted. A similar idea is applied in the event-triggered projected gradient descent algorithm \cite{YS2016}, where a sparsity projection operator is combining with the gradient descent. Although these algorithms are computationally effective, the estimation correctness can only be guaranteed under the restrict assumptions on the system structure. The satisfiability modulo theory (SMT) approach proposed in \cite{YS2017} opens up a new road to reduce the combinatorial complexity, where the computation burden is reduced by pruning the search space. Motivated by this idea, the set theory approaches are further developed in \cite{AY2017} and \cite{LA2018}. In \cite{AY2017}, less number of subset candidates induced by removing $2s$ ($s$ denotes the number of attacked sensors) sensors is used to cover the combinatorial subset candidates induced by removing $s$ sensors such that the search space is narrowed. However, the estimation performance is potentially degraded due to the remove of extra healthy measurements (as illustrated in later simulation). In \cite{LA2018}, the whole combinatorial search space is partitioned into a few of blocks such that a time varying direction descent step can be derived to compensate for the increase of Lyapunov function caused by the sparsity projection operator. However, the set partition algorithm in the off-line phrase is computational intensive. Besides these two class of approaches, some learning mechanisms have also been proposed, such as adaptive switching \cite{LW2017}, machine learning \cite{ZY2018}. However, the estimation performance cannot be guaranteed in the learning phrase of the attack mode.

In the above mentioned results, they do not consider the {\it sensor properties} associated with the contained state information. In fact, a networked sensor system usually consists of a few types of sensors \cite{SA2011,IS2005,KS2010} and the same type of sensors provides the same or equivalent measurement information. Thus, these approaches become conservative in terms of computational performance, and it is very relevant to develop an SSE scheme by taking the sensor properties into account.

This paper proposes a new SSE method which reduces the computation complexity by exploiting sensor categorization techniques, from a practical consideration. First, a concept of {\it analytic sensor types} is introduced to describe the observation characteristics of a sensor based on the equivalent class. We prove that the equivalence between two sensors can be checked in term of solving a simple linear matrix inequality (LMI). Further, we define the measurement agreement of a sensor type to assess the similarity degree of its measurement data. It is shown that, by checking the measurement agreement, the sensor type can achieve an attack detection and isolation to narrow the combinatorial search space without loss of estimation correctness. The proposed algorithm is efficient, since a networked sensor system usually consists of a few sensor types as mentioned before. Simulation comparisons are provided to illustrate high speed performance and high estimation accuracy of the algorithm compared with the existing results.


{\bf Notation:} $\mathbb{N}$ and $\mathbb{R}$ denote the set of positive natural and real numbers, respectively. The cardinality of a set $\mathbb{S}$ is denoted by $|\mathbb{S}|$ and the support of a vector $y\in \mathbb{R}^p$ is defined as supp($y$). We say that a vector $y\in\mathbb{R}^p$ is $s$-sparse if $|\mathrm{supp}(y)|\le s$. $\emptyset$ represents the empty set. For a group of sets $\mathbb{S}_i$, $i=1,\cdots,n$, $\cup_{i\in\{1,\cdots,n\}}\mathbb{S}_i=\mathbb{S}_1\cup\cdots\cup \mathbb{S}_n$.
Given $\tau$ vectors of the same dimension $y_1,\cdots,y_\tau\in\mathbb{R}^p$,
we call $y=[y_1^T,\cdots,y_\tau^T]^T\in \mathbb{R}^{\tau p}$ a block vector and each
component $y_i$ a block.
For a vector $y\in\mathbb{R}^p$, denote by $\|y\|_2$ the 2-norm of $y$ and by $\|X\|_2$ the induced 2-norm of a matrix $X\in \mathbb{R}^{m\times n}$.
For two appropriate-dimension matrices $X,Y$, denote by $(X,Y)$ the block matrix $[X^T~Y^T]^T$.
For a matrix $X\in \mathbb{R}^{m\times n}$, denote by $\mathrm{rank}(X)$ the rank of $X$, by $X^+$ the pseudo inverse matrix of $X$, and by $\sigma_M(X)$ and $\sigma_m(X)$ the maximum and minimum singular value of $X$, respectively. We also denote by $X_i\in \mathbb{R}^{1\times n}$ the $i$th row of $X$. $I_n$ represents the $n$-dimension identity matrix.

\section{Problem Formulation and Preliminaries}

Consider the following discrete-time linear dynamic system
\begin{equation}
\begin{aligned}
x(t+1)=&Ax(t)+Bu(t)+w(t)\\
y(t)=&Cx(t)+a(t)+v(t)
\end{aligned}
\end{equation}
where $x(t)\in \mathbb{R}^n$, $y(t)\in \mathbb{R}^p$ and $u(t)\in \mathbb{R}^m$ are the system state, the measured output and the control input, respectively. $w(t)\in\mathbb{R}^n$ and $v(t)\in\mathbb{R}^p$ represent bounded process and measurement noises. $a(t)\in \mathbb{R}^p$ is an $s$-sparse vector representing the attack signal injected by the adversary in different sensors and its support $\mathrm{supp}(a(t))$ is constant over time $t$. Here we only assume knowledge of an upper bound on the number of attacked sensors, we impose no constraint on how the measurements
are corrupted. The attack model is common in the SSE literature \cite{CL2015,FP2013,MS2015,HF2014,MP2017,YS2016,YS2017,AY2017,LA2018}.

 We denote by $\mathbb{S}=\{\mathbf{S}_i,i=1,\cdots,p\}$ the sensor set, and define the combinatorial set $\Sigma=\{\Gamma\subset\mathbb{S}:|\Gamma|=s\}$. Given a (block) matrix $X$ and a sensor set $\Gamma\in\Sigma$, we denote by $X_\Gamma$ the (block) matrix obtained from $X$ by removing row (block) $X_i$, $\mathbf{S}_i\in\Gamma$.

{\bf Definition 1 (Sparse observability) \cite{YS2016}.} System (1) is said to be $s$-sparse observable if for every set $\Gamma\in\Sigma$, the pair $(C_\Gamma,A)$ is observable.


{\bf Assumption 1.} System (1) is $2s$-sparse observable.

Assumption 1 is a necessary and sufficient condition for guaranteeing the existence of state estimation algorithm under $s$-sparse attack ($s<p/2$) \cite{CL2015,FP2013,MS2015,HF2014,MP2017,YS2016,YS2017,AY2017,LA2018}.

By collecting $\tau\in \mathbb{N}$ observations with $\tau\le n$, we can write the output of sensor $\mathbf{S}_i$ as follows
\begin{equation}
\begin{aligned}
\tilde Y_i(t)=\mathcal{O}_ix(t-\tau+1)+\mathbf{a}_i(t)+F_iU(t)+\Psi_i(t)
\end{aligned}
\end{equation}
where $\mathcal{O}_i=(C_i,C_iA,\cdots,C_iA^{\tau-1})$ and
$$
\begin{aligned}
\tilde Y_i(t)=&\left[\begin{matrix}
y_i(t-\tau+1)\\
y_i(t-\tau+2)\\
\vdots\\
y_i(t)\\
\end{matrix}\right],
\mathbf{a}_i(t)=\left[\begin{matrix}
a_i(t-\tau+1)\\
a_i(t-\tau+2)\\
\vdots\\
a_i(t)\\
\end{matrix}\right],\\
F_i=&\left[\begin{matrix}
0 & 0 & \cdots  &  0&  0\\
C_iB & 0 &\cdots &  0  & 0 \\
\vdots  & \ddots &\ddots & \vdots & \vdots\\
C_iA^{\tau-2}B &C_iA^{\tau-3}B & \cdots & C_iB & 0
\end{matrix}\right],\\
\Psi_i(t)=&\left[\begin{matrix}
0 & 0 & \cdots  &  0&  0\\
C_i & 0 &\cdots &  0  & 0 \\
\vdots  & \ddots &\ddots & \vdots & \vdots\\
C_iA^{\tau-2} &C_iA^{\tau-3} & \cdots & C_i& 0
\end{matrix}\right]\left[\begin{matrix}
w(t-\tau+1)\\
w(t-\tau+2)\\
\vdots\\
w(t)
\end{matrix}\right]\\
&+\left[\begin{matrix}
v_i(t-\tau+1)\\
v_i(t-\tau+2)\\
\vdots\\
v_i(t)
\end{matrix}\right],
U(t)=\left[\begin{matrix}
u(t-\tau+1)\\
u(t-\tau+2)\\
\vdots\\
u(t)\\
\end{matrix}\right].
\end{aligned}
$$

Since $U(t)$ is known, we can further simplify (2) as
\begin{equation}
Y_i(t)=\mathcal{O}_ix(t-\tau+1)+\mathbf{a}_i(t)+\Psi_i(t)
\end{equation}
where $Y_i(t)=\tilde Y_i(t)-F_iU(t)$. There exists a constant $\bar\Psi_i$ such that $\|\Psi_i(t)\|\le\bar \Psi_i$ following from the boundedness assumption on the process and measurement noise. Let $\bar\Psi=(\sum_{i=1}^p\bar\Psi_i^2)^{1/2}$.

Let $x^*\in \mathbb{R}^n$ represent the true system state. Then $Y_i=\mathcal{O}_ix^*+\mathbf{a}_i+\Psi_i$, $i=1,\cdots,p$. We dropped the time $t$ argument since the state estimation is to be solved at every time instant. Let $\mathcal{O}=(\mathcal{O}_1,\cdots,\mathcal{O}_p)$ and $Y=(Y_1,\cdots,Y_p)$. A simple exhaustive search (EX-SEARCH) algorithm is given below to derive a solution of state estimation.

\noindent{\bf Algorithm 1.} EX-SEARCH($\Sigma,\kappa$)
\hrule
\hrule
1: Initialize $\Gamma\in\Sigma$, $\hat x:=\mathcal{O}_\Gamma^+Y_\Gamma$;\\
2: {\bf while $\|Y_\Gamma-\mathcal{O}_\Gamma\hat x\|_2^2\ge\kappa\bar\Psi+\epsilon$} {\bf do}\\
3:~~~$\Sigma:=\Sigma\setminus\{\Gamma\}$;\\
4:~~~Pick $\Gamma\in\Sigma$, $\hat x:=\mathcal{O}_\Gamma^+Y_\Gamma$;\\
5: {\bf end while}\\
6: {\bf return} $\hat x$
\vspace{0.6mm}
\hrule
\hrule

{\bf Lemma 1.} Under Assumption 1, the return value of EX-SEARCH($\Sigma,\Delta_s$) remains within a small neighborhood of state $x^*$, i.e., $\|\hat x-x^*\|_2<[(\Delta_s+1)\bar\Psi+\epsilon]/\delta_{2s}$ where $\Delta_s=\max_{\Gamma\in\Sigma}\sigma_M(I-\mathcal{O}_\Gamma\mathcal{O}_\Gamma^+)$,  $\mathcal{O}_\Gamma^+=(\mathcal{O}_\Gamma^T\mathcal{O}_\Gamma)^{-1}\mathcal{O}_\Gamma^T$ is the Moore-Penrose pseduo inverse of $\mathcal{O}_\Gamma$, $\delta_{2s}=\min_{\Gamma\in\mathbb{S},|\Gamma|\le2s}\sigma_m(\mathcal{O}_\Gamma)$\footnote[1]{Under Assumption 2, one has $\delta_{2s}>0$ according to the Proposition 3.4 of \cite{YS2016}.}, and $\epsilon>0$ is the user-defined tolerance.

{\bf Proof.} By exhaustively searching $\Sigma$, there exists a set  $\Gamma\supset\mathbb{S}_{\mathrm{supp}(a)}\triangleq\{\mathbf{S}_i:i\in\mathrm{supp}(a)\}$. Thus the return $\hat x$ can take value
$$
\begin{aligned}
\hat x=&\mathcal{O}_\Gamma^+Y_\Gamma=\mathcal{O}_\Gamma^+(\mathcal{O}_\Gamma x^*+\Psi_\Gamma)\\
=&x^*+\mathcal{O}_\Gamma^+\Psi_\Gamma.
\end{aligned}
$$

Further, we have
$$
\begin{aligned}
\|Y_\Gamma-\mathcal{O}_\Gamma\hat x\|_2^2&=\|\mathcal{O}_\Gamma x^*+\Psi_\Gamma-\mathcal{O}_\Gamma(x^*+\mathcal{O}_\Gamma^+\Psi_\Gamma)\|_2^2\\
&=\|(I-\mathcal{O}_\Gamma\mathcal{O}_\Gamma^+)\Psi_\Gamma\|_2^2\le\Delta_s^2\bar\Psi^2,
\end{aligned}
$$
which guarantees that the {\bf while}-loop can be terminated and the return value $\hat x$ satisfies
$\|Y_\Gamma-\mathcal{O}_\Gamma\hat x\|_2<\Delta_s\bar\Psi+\epsilon$.

By removing the row blocks indexed by $\mathrm{supp}(a)$, one has
$$
\|\mathcal{O}_{\Gamma\cup\mathbb{S}_{\mathrm{supp}(a)}}(x^*-\hat x)+\Psi_{\Gamma\cup\mathbb{S}_{\mathrm{supp}(a)}}\|_2
<\Delta_s\bar\Psi+\epsilon
$$
which implies that $\|\hat x-x^*\|_2
<[(\Delta_s+1)\bar\Psi+\epsilon]/\delta_{2s}$. $\hfill{}$ $\blacksquare$


Algorithm 1 typically suffers from a major scalability drawback due to the combinatorial $\mathcal{O}_\Gamma$. In fact, many sensors often have the same observation characteristics (directly or equivalently measure the same state information) in CPSs \cite{SA2011,IS2005,KS2010}. From the practical consideration, the objective of this paper is to design a low-complexity SSE algorithm by narrowing the search space $\Sigma$ while guaranteeing the estimation correctness.

\section{Main results}

In this section, we consider the problem of SSE under the framework proposed in Section II. In Section III-A, a concept of analytic sensor type is introduced based on the equivalent class. In Section III-B, we provide a fast SSE algorithm via an equivalence partitioning technique.

\subsection{Analytic Sensor Type Based on Equivalent Class}

Before giving the definition of sensor type, we first introduce the concept of equivalent class.

{\bf Definition 2 (Equivalence relation and Equivalent class) \cite{HM1994}.} Given a set $\Theta$, let $\mathcal{T}:\Theta\to\Theta$ be an binary relation. If $\mathcal{T}$ satisfies:
\begin{itemize}
  \item Reflexivity: $\forall \theta\in \Theta\Rightarrow (\theta,\theta)\in\mathcal{T}$
  \item Symmetry: $(\theta_1,\theta_2)\in \mathcal{T}\wedge\theta_1\ne\theta_2\Rightarrow (\theta_2,\theta_1)\in\mathcal{T}$
  \item Transitivity: $(\theta_1,\theta_2)\in\mathcal{T},(\theta_2,\theta_3)\in\mathcal{T}\Rightarrow (\theta_1,\theta_3)\in\mathcal{T}$
\end{itemize}
then $\mathcal{T}$ is an equivalence relation defined on $\Theta$. If $(\theta_1,\theta_2)\in \mathcal{T}$, we say that $\theta_1$ is equivalent to $\theta_2$, denoting $\theta_1\overset{\mathcal{T}}{\backsim}\theta_2$. For a given $\theta\in \Theta$, we denote by $[\theta]_\mathcal{T}=\{\theta'\in \Theta:\theta'\overset{\mathcal{T}}{\backsim}\theta\}$ an equivalent class.

Let $\mathcal{T}$ be an equivalence relation defined on set $\Theta$. Then all the equivalent classes of $\mathcal{T}$ compose a partition of set $\Theta$, called equivalence partitioning \cite{HM1994} and denoted as $\mho_{\mathcal{T}}(\Theta)$.

{\bf Lemma 2.} Let $\mathcal{T}$ be the binary relation on $\mathbb{S}$ and $\mathbf{S}_i\overset{\mathcal{T}}{\backsim}\mathbf{S}_j$ if there exists a nonsingular matrix $T_{ij}\in \mathbb{R}^{\tau\times\tau}$ such that $T_{ij}\mathcal{O}_j=\mathcal{O}_i$. Then $\mathcal{T}$ is an equivalence relation on set $\mathbb{S}$.

{\bf Proof.} We show that three properties in Definition 2 hold.

Reflexivity. Noting that the fact $I_\tau\mathcal{O}_i=\mathcal{O}_i$ for any $i\in\{1,2,\cdots,p\}$, then $\mathbf{S}_i\overset{\mathcal{T}}{\backsim}\mathbf{S}_i$.

Symmetry. Since $T_{ij}$ is nonsingular, then $T_{ij}\mathcal{O}_j=\mathcal{O}_i\Leftrightarrow T_{ji}^{-1}\mathcal{O}_i=\mathcal{O}_j$, i.e., $\mathbf{S}_i\overset{\mathcal{T}}{\backsim}\mathbf{S}_j\Leftrightarrow \mathbf{S}_j\overset{\mathcal{T}}{\backsim}\mathbf{S}_i$.

Transitivity. Assume that $\mathbf{S}_i\overset{\mathcal{T}}{\backsim}\mathbf{S}_j$ and $\mathbf{S}_j\overset{\mathcal{T}}{\backsim}\mathbf{S}_k$. Then there exist nonsingular matrices $T_{ij}$ and $T_{jk}$ such that $T_{ij}\mathcal{O}_j=\mathcal{O}_i$ and $T_{jk}\mathcal{O}_k=\mathcal{O}_j$. We have $T_{ij}T_{jk}\mathcal{O}_k=\mathcal{O}_i$, which means that $\mathbf{S}_i\overset{\mathcal{T}}{\backsim}\mathbf{S}_k$ by choosing $T_{ik}=T_{ij}T_{jk}$. $\hfill{}$ $\blacksquare$

Based on Lemma 2, the set of all the sensors is partitioned into different equivalent classes and we define every equivalent class as an {\bf analytic sensor type}. In fact, according to the definition of $\mathcal{T}$, if $\mathbf{S}_i\overset{\mathcal{T}}{\backsim}\mathbf{S}_j$ then under attack- and noise-free case, we have
\begin{align}
Y_i&=\mathcal{O}_ix^*,\\
T_{ij}Y_j&=T_{ij}\mathcal{O}_jx^*=\mathcal{O}_ix^*,
\end{align}
which means that sensors $\mathbf{S}_i$ and $\mathbf{S}_j$ substantially monitor the same state information of system (1), which is called observation equivalence.

{\bf Remark 1.} The ``analytic'' is used to distinguish the concept of {\it hardware sensor type}  introduced in \cite{SA2011}. This is motivated by the analytic redundancy and hardware redundancy in fault detection and isolation literature. In \cite{SA2011}, sensors $\mathbf{S}_i$
and $\mathbf{S}_j$ are in the same type only if $C_i=C_j$. Instrinsically, this definition describes the sensor hardware redundancy because the same type of sensors is used to monitor the same one system state (or linear combination of same system states). In this paper, a more general concept of sensor type is proposed, where two sensors which measure different two state entries may also belong to the
same one analytic sensor type (see the following Examples 1 and 2). This concept clearly characterizes the connection and difference between two sensors in the sense of observability. In the following context, without confusing, the ``analytic sensor type'' is simply written as ``sensor type''.

The following examples show that the proposed concept of analytic sensor type is more relaxed than the hardware type in \cite{SA2011}. As a result, it provides a more compact partition than the hardware sensor type.

{\bf Example 1.} Consider the F-16 short period dynamics \cite{BL2003}, which is modeled as (1) with
$$
\begin{aligned}
&A=\left[\begin{matrix}
9.0649\times 10^{-1} & 8.1601\times 10^{-2} & -5.0128\times 10^{-4}\\
7.4135\times 10^{-2} & 9.0121\times 10^{-1}  & -7.0423\times 10^{-3}\\
0 & 0 & 1.3266\times 10^{-1}
\end{matrix}\right],\\
&C_1=\left[\begin{matrix}
1&  0 & 0
\end{matrix}\right]~\mathrm{and}~ C_2=\left[\begin{matrix}
0&  1 & 0
\end{matrix}\right].
\end{aligned}
$$
Two sensors monitor the angle of attack and pitch rate, and then belong to different hardware types \cite{SA2011}. However, it is verified that these two sensors belong to the same analytic type based on the equivalent relation $\mathcal{T}$ (simply choose $T_{12}=\mathcal{O}_1\mathcal{O}_2^{-1}$).

{\bf Example 2.} Let us consider a B747-100/200 aircraft borrowed from NASA \cite{HC1970}. By choosing the sampling period $h=0.1$s, the discrete-time linear model is obtained with
$$
A=\left[\begin{matrix}
0.9337  &  0.0000  & -0.0099 &  -0.0000  \\
-0.0845 &   0.9994 &  -0.4537&   -0.9791 \\
0.0944  & -0.0001  &  0.9440 &   0.0000 \\
0.0967  &  0.0000  & -0.0005 &   1.0000
\end{matrix}\right]
$$
and all states containing pitch rate, true airspeed, angle of attack and pitch angle are measurable, i.e., $C=I_4$. It can be verified that all four sensors belong to the same one analytic type by letting $T_{ij}=\mathcal{O}_i\mathcal{O}_j^{-1}$, $i,j=1,2,3,4$.

%

\subsection{SSE Algorithm Based on Equivalence Partitioning}
In this section, a subset extraction algorithm is first given to find the equivalence partition of $\mathbb{S}$. Its intuitive idea is to extract the equivalent class of sensor $\mathbf{S}_i$, denoted by $[\mathbf{S}_i]_{\mathcal{T}}$, as a block of the partition and exclude it from $\mathbb{S}$, one-by-one, until the remaining set is empty. Then the intuition can be simplified into the following alternation:
$$
\begin{aligned}
&[\mathbf{S}_i]_{\mathcal{T}}:=\text{ECLA}\cdot \text{SOLVE}(\mathbf{S}_i)\\
&\mho_{\mathcal{T}}:=\mho_{\mathcal{T}}\cup[\mathbf{S}_i]_{\mathcal{T}}
\end{aligned}
$$
where $\mathbf{S}_i\in \mathbb{S}\setminus\bigcup_{[\mathbf{S}_j]_\mathcal{T}\in\mho_\mathcal{T}}[\mathbf{S}_j]_\mathcal{T}$ and $\text{ECLA}\cdot \text{SOLVE}(\mathbf{S}_i)$ is the operator of solving the equivalent class of sensor $\mathbf{S}_i$. Formalize these ideas to the following algorithm.

\noindent{\bf Algorithm 2.} Equivalence Partitioning (EP$\cdot$SOLVE$_\mathcal{T}(\mathbb{S})$)
\hrule
\hrule
1: Initialize $\mathbb{S}_0:=\mathbb{S}$, $\mho_\mathcal{T}:=\emptyset$;\\
~2: {\bf while} $\mathbb{S}_0\ne\emptyset$ {\bf do}\\
~3:~~~~Pick $\mathbf{S}_i\in\mathbb{S}_0$;\\
~4:~~~~$[\mathbf{S}_i]_{\mathcal{T}}:=\{\mathbf{S}_i\}$;\\
~5:~~~~{\bf for} $\mathbf{S}_j\in \mathbb{S}_0\setminus\mathbb{S}_{\mathrm{equ}}^i$ {\bf do}\\
~6:~~~~~~~{\bf if} $\mathbf{S}_i\overset{\mathcal{T}}{\backsim}\mathbf{S}_j$ {\bf then}\\
~7:~~~~~~~~~~$[\mathbf{S}_i]_{\mathcal{T}}:=[\mathbf{S}_i]_{\mathcal{T}}\cup\{\mathbf{S}_j\}$;\\
~8:~~~~~~~{\bf end if}\\
~9:~~~~{\bf end for}\\
10:~~~~$\mho_\mathcal{T}:=\mho_\mathcal{T}\cup\{[\mathbf{S}_i]_{\mathcal{T}}\}$;\\
11:~~~~$\mathbb{S}_0:=\mathbb{S}_0\setminus\{[\mathbf{S}_i]_{\mathcal{T}}\}$;\\
12: {\bf end while}\\
13: {\bf return} $\mho_{\mathcal{T}}$
\hrule
\hrule
\vspace{1mm}

Following the above line, the equivalence partition $\mho_\mathcal{T}$ can be obtained by performing EP$\cdot$SOLVE$_\mathcal{T}(\mathbb{S})$. Note that the {\bf if}-criterion  that $\mathbf{S}_i\overset{\mathcal{T}}{\backsim}\mathbf{S}_j$ in line 6 of Algorithm 2 is not directly exercisable. The following lemma provides an LMI method to check the {\bf if}-criterion.

{\bf Lemma 3.} Given two sensors $\mathbf{S}_i,\mathbf{S}_j\in \mathbb{S}$, then $\mathbf{S}_i\overset{\mathcal{T}}{\backsim}\mathbf{S}_j$ if and only if $\mathrm{rank}(\mathcal{O}_i)=\mathrm{rank}(\mathcal{O}_j)$ and there exists a nonsingular matrix $T_{ij}\in \mathbb{R}^{\tau\times\tau}$ such that the following LMI holds with $\varepsilon\approx 0$
\begin{align}
\left[\begin{matrix}
-\varepsilon I& *\\
T_{ij}\mathcal{O}_j-\mathcal{O}_i &-\varepsilon I
\end{matrix}\right]<0
\end{align}

{\bf Proof.} Necessity. Since $\mathbf{S}_i\overset{\mathcal{T}}{\backsim}\mathbf{S}_j$, there exists a nonsingular matrix $T_{ij}$ such that $T_{ij}\mathcal{O}_j=\mathcal{O}_i$. Since $T_{ij}$ is nonsingular then $\mathrm{rank}(\mathcal{O}_i)=\mathrm{rank}(\mathcal{O}_j)$ and $T_{ij}$ is a solution of LMI (6).

Sufficiency. LMI (6) guarantees that there exists $T'_{ij}$ such that $T_{ij}'\mathcal{O}_j=\mathcal{O}_i$. Next we show that there must exist a nonsingular matrix $T_{ij}$ such that $T_{ij}\mathcal{O}_j=\mathcal{O}_i$ under the condition $\mathrm{rank}(\mathcal{O}_i)=\mathrm{rank}(\mathcal{O}_j)$.

According to the linear algebra theory, there exist nonsingular matrices $T_i$ and $T_j$ such that
$$
T_i\mathcal{O}_i=\left[\begin{matrix}
\mathcal{O}_i^1\\
0
\end{matrix}\right],~T_j\mathcal{O}_j=\left[\begin{matrix}
\mathcal{O}_j^1\\
0
\end{matrix}\right]
$$
where $\mathcal{O}_i^1$ and $\mathcal{O}_j^1$ are of full row rank and $\mathrm{rank}(\mathcal{O}_i^1)=\mathrm{rank}(\mathcal{O}_j^1)$.
With regard to $T_{ij}'\mathcal{O}_j=\mathcal{O}_i$, one has
$$
T_iT_{ij}'T_j^{-1}\left[\begin{matrix}
\mathcal{O}_j^1\\
0
\end{matrix}\right]=\left[\begin{matrix}
\mathcal{O}_i^1\\
0
\end{matrix}\right]
$$
and matrix $T_iT_{ij}'T_j^{-1}$ has the form
$$T_iT_{ij}'T_j^{-1}=\left[\begin{matrix}
\Lambda& *\\
0 &*'
\end{matrix}\right]
$$
where $\Lambda$ is a nonsingular matrix such that $\Lambda\mathcal{O}_j^1=\mathcal{O}_i^1$ and $*$ and $*'$ represent any certain two matrices.

Choose $*'$ to be nonsingular and define
$$
T_{ij}\overset\vartriangle=T_j^{-1}\left[\begin{matrix}
\Lambda& *\\
0 &*'
\end{matrix}\right]T_i.
$$
Then $T_{ij}$ is nonsingular and satisfies $T_{ij}\mathcal{O}_j=\mathcal{O}_i$. $\hfill{} \blacksquare$

From Lemma 3, to test the observation equivalence of two sensors by LMI (6), we first should make an equal rank partitioning for set $\mathbb{S}$, and then perform EP$\cdot$SOLVE$_\mathcal{T}$ for each equal rank block. Let $\mathcal{T}_r$ be the equal rank relation on $\mathbb{S}$ such that $\mathbf{S}_i\overset{\mathcal{T}_r}{\backsim}\mathbf{S}_j$ if $\mathrm{rank}(\mathcal{O}_j)=\mathrm{rank}(\mathcal{O}_i)$. It is easy to be verified that $\mathcal{T}_r$ is also an equivalence relation on set $\mathbb{S}$.

Thus, we derive the final equivalence partitioning algorithm.

\noindent{\bf Algorithm 3.} Double-Equivalence Partitioning
\hrule
\hrule
1: Initialize $\mathbb{S}_0:=\mathbb{S}$, $\mho_\mathcal{T}:=\emptyset$;\\
2: $\mho_{\mathcal{T}_r}:=$EP$\cdot$SOLVE$_{\mathcal{T}_r}(\mathbb{S}_0)$;\\
3:~~~{\bf for} $\mathbb{S}_{\mathrm{temp}}\in \mho_{\mathcal{T}_r}$ {\bf do}\\
4:~~~~~~$\mho_\mathcal{T}:=\mho_\mathcal{T}\cup$EP$\cdot$SOLVE$_\mathcal{T}(\mathbb{S}_{\mathrm{temp}})$;\\
5:~~~{\bf end for}\\
6: {\bf return} $\mho_\mathcal{T}$
\hrule
\hrule
\vspace{1mm}

Using the Double-Equivalence Partitioning (termed DE-Partitioning), all the sensors are categorized into a few types. Given a sensor $\mathbf{S}_j\in [\mathbf{S}_i]_{\mathcal{T}}$, its output equation (3) can be further rewritten as
\begin{equation}
\hat Y_j=\mathcal{O}_ix^*+\hat{\mathbf{a}}_j+\hat\Psi_j,~
\end{equation}
where $\hat Y_j=T_{ij}Y_j$,  $\hat{\mathbf{a}}_j=T_{ij}\mathbf{a}_j$, $\hat\Psi_j=T_{ij}\Psi_j$ and $T_{ij}$ can be obtained by solving LMI (6).

From (7), when sensor type $[\mathbf{S}_i]_{\mathcal{T}}$ is attack-free, i.e., $\hat{\mathbf{a}}_j=0$, the outputs of any two sensors in this type are very near in the presence of a sufficiently small disturbance $\hat\Psi_j$. Oppositely, when type $[\mathbf{S}_i]_{\mathcal{T}}$ is attacked, the sensor measurements  may be markedly different, especially between attacked and attack-free sensors. To quantify the similarity degree of measurement data in a sensor type, we introduce a concept of measurement agreement (M-Agreement) for the sensor types.

{\bf Definition 3 (Median-Agreement).} For a given sensor type $[\mathbf{S}_i]_{\mathcal{T}}\in\mho_{\mathcal{T}}^\diamond$, if $\|\hat Y_j-\check Y\|\le 2M_T\bar\Psi$ holds for any $\mathbf{S}_j\in [\mathbf{S}_i]_{\mathcal{T}}$ where $M_T=\max_{\mathbf{S}_j\in[\mathbf{S}_i]_{\mathcal{T}},[\mathbf{S}_i]_\mathcal{T}\in\mho^\star_\mathcal{T}\cup\mho^\diamond_\mathcal{T}}\|T_{ij}\|$,
then we say that sensor type $[\mathbf{S}_i]_{\mathcal{T}}$ is Median-Agreeable, where $\check Y$ is called vector median whose $m$-th element is the median of the sequence consisting of the $m$-th element of $\hat Y_j$, $\mathbf{S}_j\in [\mathbf{S}_i]_{\mathcal{T}}$.

It is important to note that, a bigger sensor type can generally provide more attack location information. Particularly, the block $[\mathbf{S}_i]_{\mathcal{T}}\in\mho_{\mathcal{T}}$ with $|[\mathbf{S}_i]_{\mathcal{T}}|=1$ is trivial and contains no attack information. Also, if there exists the sensor type that contains at least $2s+1$ sensors, then the number of attack-free measurements is more than that of attacked measurements. As a result, the vector median given in Definition 3 can be regarded as an attack-free measurement. Based on this,  we consider two subsets of $\mho_{\mathcal{T}}$: $\mho_{\mathcal{T}}^\star=\{[\mathbf{S}_i]_{\mathcal{T}}\in\mho_{\mathcal{T}}:2\le|[\mathbf{S}_i]_{\mathcal{T}}|<2s+1\}$ and $\mho_{\mathcal{T}}^\diamond=\{[\mathbf{S}_i]_{\mathcal{T}}\in\mho_{\mathcal{T}}:|[\mathbf{S}_i]_{\mathcal{T}}|\ge2s+1\}$.

We denote by $\mho_{\mathcal{T}}^+$ the subset of $\mho_{\mathcal{T}}^\star$ consisting of Mean-Agreeable sensor types, and denote $\mho_{\mathcal{T}}^-=\mho_{\mathcal{T}}^\star\setminus\mho_{\mathcal{T}}^+$. Further, we define $\mho_{\mathcal{T}}^{\bar s}=\{[\mathbf{S}]_{\mathcal{T}}\in\mho_{\mathcal{T}}^+:|[\mathbf{S}]_{\mathcal{T}}|>s\}$ and $\mho_{\mathcal{T}}^{\underline s}=\mho_{\mathcal{T}}^+\setminus\mho_{\mathcal{T}}^{\bar s}$.

For given an equivalent class $[\mathbf{S}_i]_{\mathcal{T}}\in\mho_{\mathcal{T}}^\diamond$, we define $[\mathbf{S}_i]_{\mathcal{T}}^\ominus\triangleq\{\mathbf{S}_j\in [\mathbf{S}_i]_{\mathcal{T}}:\|\hat Y_j-\check Y\|>2M_T\bar\Psi\}$, and denote $[\mathbf{S}_i]_{\mathcal{T}}^\oplus=[\mathbf{S}_i]_{\mathcal{T}}\setminus[\mathbf{S}_i]_{\mathcal{T}}^\ominus$. If $[\mathbf{S}_i]_{\mathcal{T}}^\ominus=\emptyset$, then $[\mathbf{S}_i]_{\mathcal{T}}$ is Median-Agreeable.

Based on the equivalence partitioning and the M-Agreement of each block, the attack location information can be derived. This is shown in the following propositions.

{\bf Proposition 1.} Consider a sensor type $[\mathbf{S}_i]_{\mathcal{T}}\in\mho_{\mathcal{T}}^\star$ and assume that each attack signal $\hat{\mathbf{a}}_j$, $\mathbf{S}_j\in[\mathbf{S}_i]_{\mathcal{T}}$ satisfies
$\|\hat{\mathbf{a}}_j\|>4M_T\bar\Psi$.
Then, the following statements hold
\begin{description}
  \item[i)]If $[\mathbf{S}_i]_{\mathcal{T}}\in\mho_{\mathcal{T}}^-$, then $[\mathbf{S}_i]_{\mathcal{T}}$ must contain attacked sensors, i.e., $\mathbb{S}_{\mathrm{supp}(a)}\cap[\mathbf{S}_i]_{\mathcal{T}}\ne\emptyset$;
  \item[ii)] If $[\mathbf{S}_i]_{\mathcal{T}}\in\mho_{\mathcal{T}}^{\bar s}$, then all the sensors in $[\mathbf{S}_i]_{\mathcal{T}}$ are not attacked, i.e., $\mathbb{S}_{\mathrm{supp}(a)}\cap[\mathbf{S}_i]_{\mathcal{T}}=\emptyset$;
  \item[iii)]If $[\mathbf{S}_i]_{\mathcal{T}}\in\mho_{\mathcal{T}}^{\underline s}$, then all the sensors in $[\mathbf{S}_i]_{\mathcal{T}}$ are not attacked or all are attacked, i.e., $\mathbb{S}_{\mathrm{supp}(a)}\cap[\mathbf{S}_i]_{\mathcal{T}}=\emptyset$ or $
      \mathbb{S}_{\mathrm{supp}(a)}\supset[\mathbf{S}_i]_{\mathcal{T}}$.
\end{description}

{\bf Proof.} Next we show these three statements by contradiction.

i) Supposing that all the sensors in $[\mathbf{S}_i]_{\mathcal{T}}$ are not attacked, i.e., $\mathbb{S}_{\mathrm{supp}(a)}\cap[\mathbf{S}_i]_{\mathcal{T}}=\emptyset$, one has
$$
\begin{aligned}
\hat Y_j&=T_{ij}\mathcal{O}_jx^*+T_{ij}\Psi_j=\mathcal{O}_ix^*+T_{ij}\Psi_j, \forall \mathbf{S}_j\in[\mathbf{S}_i]_{\mathcal{T}}
\end{aligned}
$$
and there exists a vector $\check\Psi$ satisfying $\|\check\Psi\|\le M_T\bar\Psi$ such that
$$
\check Y=\mathcal{O}_ix^*+\check\Psi.
$$

It follows that
$\left\|\hat Y_j-\tilde Y\right\|\le2M_T\bar\Psi$
which means that $[\mathbf{S}_i]_{\mathcal{T}}\in\mho_{\mathcal{T}}^+$, a contradiction.

ii) Supposing that $\mathbb{S}_{\mathrm{supp}(a)}\cap[\mathbf{S}_i]_{\mathcal{T}}\ne\emptyset$, then $[\mathbf{S}_i]_{\mathcal{T}}$ must simultaneously contain attacked sensors and attack-free sensors since $|[\mathbf{S}_i]_{\mathcal{T}}|>s$. Assume that $\mathbf{S}_j$ is attacked but $\mathbf{S}_k$ is not. Then one has
$$
\begin{aligned}
\hat Y_j&=\mathcal{O}_ix^*+\hat{\mathbf{a}}_j+\hat\Psi_j,\\
\hat Y_k&=\mathcal{O}_ix^*+\hat\Psi_k.
\end{aligned}
$$
Since $[\mathbf{S}_i]_{\mathcal{T}}\in \mho_{\mathcal{T}}^+$, then we have
$$
\begin{aligned}
\|\hat{\mathbf{a}}_j\|-2M_T\bar\Psi&\le\|\hat Y_j-\hat Y_k\|=\|\hat{\mathbf{a}}_j+\hat\Psi_j-\hat\Psi_k\|\\
&\le\|\hat Y_j-\check Y\|+\|\check Y-\hat Y_k\|\le2M_T\bar\Psi
\end{aligned}
$$
which implies that $\|\hat{\mathbf{a}}_j\|\le4M_T\bar\Psi$, a contradiction.

iii). Suppose that a part of sensors in $[\mathbf{S}_i]_{\mathcal{T}}\in\mho_{\mathcal{T}}^{\underline s}$ are attacked but the other part are not. The situation is the exactly same as ii) and the proof can be complete.  $\hfill{}$ $\blacksquare$

{\bf Proposition 2.} Consider a sensor type $[\mathbf{S}_i]_{\mathcal{T}}\in\mho_{\mathcal{T}}^\diamond$ and assume that each attack signal $\hat{\mathbf{a}}_j$, $\mathbf{S}_j\in[\mathbf{S}_i]_{\mathcal{T}}$ satisfies $\|\hat{\mathbf{a}}_j\|>4M_T\bar\Psi.$
Then sensor $\mathbf{S}_j$ is attacked if and only if $\|\hat Y_j-\check Y\|>2M_T\bar\Psi$, i.e., $[\mathbf{S}_i]_{\mathcal{T}}^\ominus\subset\mathbb{S}_{\mathrm{supp}(a)}$ and
$[\mathbf{S}_i]_{\mathcal{T}}^\oplus\cap\mathbb{S}_{\mathrm{supp}(a)}=\emptyset$.

{\bf Proof.} According to the definition of $\check Y$ in Definition 3 and $|\mathrm{supp}(a)|\le s$, there exists a vector $\check\Psi$ satisfying $\|\check\Psi\|\le M_T\bar\Psi$ such that
$$
\check Y=\mathcal{O}_ix+\check\Psi.
$$
Under the condition that $\|\hat{\mathbf{a}}_j\|>4M_T\bar\Psi$, it is easily shown that sensor $\mathbf{S}_j$ is attacked if and only if $\|\hat Y_j-\check Y\|>2M_T\bar\Psi$. The next proof is similar to that of Proposition 1 and omitted here.$\hfill{} \blacksquare$

In fact, the check of M-Agreement of a sensor type serves as an attack detection and isolation mechanism. It can provide the attack location information and prune the search space.

{\bf Theorem 1.} Under Assumption 1, if each attack signal $\hat{\mathbf{a}}_j$, $\mathbf{S}_j\in\mho_{\mathcal{T}}^\star$ in (7) satisfies
$
\|\hat{\mathbf{a}}_j\|>4M_T\bar\Psi,$ then the return value of EX-SEARCH($\Sigma_{\mathcal{T}},\Delta_s$) remains within a neighborhood of $x^*$, i.e., $\|\hat x-x^*\|_2\le[(\Delta_s+1)\bar\Psi+\epsilon]/\delta_{2s}$,
where
$$
\begin{aligned}
\Sigma_{\mathcal{T}}&=
\Sigma\setminus(\Sigma_{\mathcal{T}}^-\cup\Sigma_{\mathcal{T}}^{\bar s}\Sigma_{\mathcal{T}}^{\underline s}\cup\Sigma_{\mathcal{T}}^\diamond)\\
\Sigma_{\mathcal{T}}^-&=\cup_{[\mathbf{S}]_{\mathcal{T}}\in\mho^-_{\mathcal{T}}}\{\Gamma\in\Sigma:\Gamma\cap[\mathbf{S}]_{\mathcal{T}}=\emptyset\},\\
\Sigma_{\mathcal{T}}^{\bar s}&=\cup_{[\mathbf{S}]_{\mathcal{T}}\in\mho^{\bar s}_{\mathcal{T}}}\{\Gamma\in\Sigma:\Gamma\cap[\mathbf{S}]_{\mathcal{T}}\ne\emptyset\},\\
\Sigma_{\mathcal{T}}^{\underline s}&=\cup_{[\mathbf{S}]_{\mathcal{T}}\in\mho^{\underline s}_{\mathcal{T}}}\{\Gamma\in\Sigma:\emptyset\varsubsetneqq(\Gamma\cap[\mathbf{S}]_{\mathcal{T}})\varsubsetneqq [\mathbf{S}]_{\mathcal{T}}\},\\
\Sigma_{\mathcal{T}}^\diamond&=\cup_{[\mathbf{S}]_{\mathcal{T}}\in\mho_{\mathcal{T}}}\{\Gamma\in\Sigma:\Gamma\nsupseteq[\mathbf{S}]_{\mathcal{T}}^\ominus~\mathrm{or}~ [\mathbf{S}]_{\mathcal{T}}^\oplus\cap\Gamma\ne\emptyset\}.
\end{aligned}
$$

{\bf Proof.} From Propositions 1 and 2, the attacked sensor set satisfies $\mathbb{S}_{\mathrm{supp}(a)}\notin \Sigma_{\mathcal{T}}^-\cup\Sigma_{\mathcal{T}}^{\bar s}\cup\Sigma_{\mathcal{T}}^{\underline s}\cup\Sigma_{\mathcal{T}}^\diamond$. Therefore, there exists $\Gamma\in\Sigma_{\mathcal{T}}$ such that $\|Y_\Gamma-\mathcal{O}_\Gamma\hat x\|_2^2\le\Delta_s\bar\Psi^2$. Then the proof is complete by following Lemma 1. $\hfill{}$ $\blacksquare$

Theorem 1 characterizes the class of detectable attack signals, i.e, $\|\hat{\mathbf{a}}_j\|>4M_T\bar\Psi$. However, a crafty attacker may ingeniously inject the attack signals (called undetectable attacks) which are not detected by the proposed M-Agreement, yet increase the estimation error (Under this case, a sensor type still keeps M-Agreeable despite being attacked). The following result evaluates the estimation error in the presence of undetectable attacks.

{\bf Theorem 2.} Under Assumption 1, the return of EX-SEARCH ($\Sigma_{\mathcal{T}},\eta\Delta_s$) remains within a neighborhood of $x^*$, i.e., $\|\hat x-x^*\|_2\le[(\eta\Delta_s+1)\bar\Psi+\epsilon]/\delta_{2s}$, where $\eta=\sqrt{1+16M_T^2m_T^2}$  and $m_T=\min_{j\in [\mathbf{S}_i]_\mathcal{T},[\mathbf{S}_i]_\mathcal{T}\in\mho^\star_\mathcal{T}\cup \mho^\diamond_\mathcal{T}}\|T_{ji}\|$.

{\bf Proof.} We first show there exists a set $\Gamma\in \Sigma_\mathcal{T}$ such that
\begin{align}
\|Y_\Gamma-\mathcal{O}_\Gamma \hat x\|_2
\le\eta\Delta_s\bar\Psi.
\end{align}
with two scenarios.

\begin{enumerate}
  \item When $\|\hat{\mathbf{a}}_j\|>4M_T\bar\Psi$, from Theorem 1 there exists a set $\Gamma$ satisfying $\mathrm{supp}(a)\subset\Gamma\in\Sigma_\mathcal{T}$ such that (8) holds.
  \item When $\|\hat{\mathbf{a}}_j\|\le4M_T\bar\Psi$, an attacked sensor type may still be M-Agreeable. Under this case, the sets $\Gamma$'s satisfying $\mathrm{supp}(a)\subset\Gamma\in\Sigma_\mathcal{T}$ may be deleted from $\Sigma_\mathcal{T}$ but another set $\Gamma_*$ that satisfies $\|\hat{\mathbf{a}}_j\|\le4M_T\bar\Psi$, $\mathbf{S}_j\in \Gamma_*$ can be contained in $\Sigma_{\mathcal{T}}$. By exhaustively searching $\Sigma_{\mathcal{T}}$, the return $\hat x$ can take value
$$
\begin{aligned}
\hat x=&\mathcal{O}_{\Gamma_*}^+Y_{\Gamma_*}=\mathcal{O}_{\Gamma_*}^+(\mathcal{O}_{\Gamma_*} x^*+\mathbf{a}_{\Gamma_*}+\Psi_{\Gamma_*})\\
=&x^*+\mathcal{O}_{\Gamma_*}^+\mathbf{a}_{\Gamma_*}+\mathcal{O}_{\Gamma_*}^+\Psi_{\Gamma_*}.
\end{aligned}
$$
Further we have
$$
\begin{aligned}
\|Y_{\Gamma_*}-\mathcal{O}_{\Gamma_*} \hat x\|_2^2\le&\|(I-\mathcal{O}_{\Gamma_*}\mathcal{O}_{\Gamma_*}^+)\Psi_{\Gamma_*}\|_2^2\\
&+\|(I-\mathcal{O}_{\Gamma_*}\mathcal{O}_{\Gamma_*}^+)\mathbf{a}_{\Gamma_*}\|_2^2\\
\le&(1+16M_T^2m_T^2)\Delta_s^2\bar\Psi^2.
\end{aligned}
$$
\end{enumerate}
Therefore, the condition (8) holds and the {\bf while}-loop can be terminated and the return value $\hat x$ satisfies $\|Y_\Gamma-\mathcal{O}_\Gamma\hat x\|_2^2<\eta\Delta_s\bar\Psi+\epsilon$. By removing the row blocks indexed by $\mathrm{supp}(a)$, one has
$$
\|\mathcal{O}_{\Gamma\cup\mathrm{supp}(a)}(x^*-\hat x)+\Psi_{\Gamma\cup\mathrm{supp}(a)}\|_2
<\eta\Delta_s\bar\Psi+\epsilon
$$
which yields that $\|\hat x-x^*\|_2
<[(\eta\Delta_s+1)\bar\Psi+\epsilon]/\delta_{2s}$. $\hfill{}$ $\blacksquare$

{\bf Corollary 1.} If system (1) is $(p-1)$-sparse observable, then the return value of EX-SEARCH($[\mathbf{S}_1]_{\mathcal{T}}^\oplus,\eta'\Delta_s$) remains within a neighborhood of $x^*$, i.e., $\|\hat x-x^*\|_2\le[(\eta'\Delta_s+1)\bar\Psi+\epsilon]/\delta_{2s}$, where $\eta'=\sqrt{1+16M_T^2m_T^2}$.

{\bf Proof.} Since system (1) is $(p-1)$-sparse observable, then $\mathcal{O}_i$, $i=1,\cdots,p$ are nonsingular, and resultantly all the sensors belong to one sensor type which can be represented by $[\mathbf{S}_1]_{\mathcal{T}}$. Also, $[\mathbf{S}_1]_{\mathcal{T}}\in \mho_{\mathcal{T}}^\diamond$. From Proposition 2 and following the proof of Theorems 1 and 2, the proof is complete. $\hfill{}$ $\blacksquare$

Corollary 1 indicates that if each single matrix pair $(C_i,A)$ is observable, then the size of search space is only one. This directly illustrates the main advantage of the proposed SSE algorithm compared with the existing \cite{YS2017,AY2017,LA2018}. Note that many practical systems satisfy such property, such as F-16 short period dynamics and B747-100/200 aircraft given in the above Examples 1 and 2.

Now, we summarize the final SSE algorithm.

\noindent{\bf Algorithm 4.} Fast Secure State Estimation (FSSE)
\hrule
\hrule
1: {\it Offline}:\\
2:~~~~Perform DE-Partitioning Algorithm;\\
3: {\it Online:} {\bf Input} $Y(t)$\\
4:~~~~Check M-Agreement of sensor types;\\
5:~~~~Run EX-SEARCH($\Sigma_{\mathcal{T}},\eta\Delta_s$);
\hrule
\hrule
\vspace{2mm}


{\bf Remark 2.} For a sensor type $[\mathbf{S}]_{\mathcal{T}}\in\mho^\star_{\mathcal{T}}$, the online check of the Median-Agreement needs to calculate the median. Recalling that the median can be computed with complexity $\mathrm{O}(|[\mathbf{S}]_{\mathcal{T}}|)$ \cite{W}, thus the computation complexity is linear.
It is well known that the calculation of the mean is more flexible than that of the median. Hence, we can introduce another concept of M-Agreement, {\it Mean-Agreement} (check $\|\hat Y_j-\tilde Y\|\le (1+M_T)\bar\Psi$ where $\tilde Y=\sum_{\mathbf{S}_l\in[\mathbf{S}]_{\mathcal{T}}}\hat Y_l/|[\mathbf{S}]_{\mathcal{T}}|$ in Definition 3), to assess the measurement agreement of $\mho_{\mathcal{T}}^\star$ in order to reduce the computation burden.



The way of reducing combinatorial complexity, cutting search space, has been adopted in \cite{YS2017,AY2017,LA2018}. In \cite{YS2017}, under the condition of $2s$-sparse observability, a certificate based on conflicting sensor sets whose cardinality is $p-2s+1$ can be designed to provide the attack information and narrow the search scope. This algorithm relies on a heuristic idea for deciding search order from all possible combinations and the worst-case size of basic search space is still $\mathcal{C}_p^{p-2s+1}$. Comparatively,
the works in \cite{AY2017}, \cite{LA2018} and this paper provide the set theory approaches to reduce the basic search space.

In \cite{AY2017}, the search space can be reduced by finding the minimal cover $\Omega$ from $\Sigma_{2s}$ of set $\Sigma$, i.e., find a minimal set $\Omega\subset\Sigma_{2s}$ such that for any set $\Gamma\in \Sigma$, there exists a set $\Gamma_{2s}\in\Omega$ such that $\Gamma\subset\Gamma_{2s}$, where $\Sigma_{2s}=\{\Gamma\subset\mathbb{S}:|\Gamma|=2s\}$. Thus, the search space becomes $\Omega$. It has been shown in \cite{AY2017} that the set cover based approach can prune at least half of search space. Nevertheless, this approach does not consider the sensor characteristics and thus it is still conservative when the networked sensor systems consist of a few types of sensors. Besides, the {\it main drawback} of the SSE approach \cite{AY2017} is that additional $s$ healthy sensor measurements have to be removed such that the estimation accuracy is degraded in the presence of measurement noise. Comparatively, the proposed SSE approach can narrow the search space without loss of estimation performance in sense that all healthy sensors are used for state estimation.

In \cite{LA2018}, the overall search space $\Sigma$ can be partitioned into $m (m\le \mathcal{C}_p^s)$ subsets $\Sigma_i$, $i=1,\cdots,m$ according to the certificate that for each subset $\Sigma_i$, a common matrix $R_i$ can be found such that
$
3/4I<R_iQ_\Gamma<I,~\forall \Gamma\in \Sigma_i
$
with $Q_\Gamma=[\mathcal{O},I_\Gamma^T]$. In this paper, the proposed approach exploits sensor characteristics to construct the local attack detection mechanism such that the search space is sufficiently narrowed. It can be seen that, the ideas and ways to narrowing search space are different. Moreover, in off-line design stage the proposed approach has less computation complexity. The set partitioning based approach \cite{LA2018} requires solving $\sum_{i=1}^{\mathcal{C}_{p-s}^s}i!$ LMIs with $(n+2s\tau)$ dimensions at the worst-case case, and each LMI contains $\tau p(n+2s\tau)$ variables. Comparatively, the proposed FSSE Algorithm requires solving $p(p-1)/2$ LMIs with $(n+\tau)$ dimensions and each LMI contains $\tau^2$ variables.

From the above analysis, the approaches in \cite{YS2017,AY2017,LA2018} and this paper narrow the search space in {\it different fashions} (cooperation of these algorithms is possible) and the techniques adopted are also fundamentally different. In practical applications, these four approaches should be appropriately selected according to the specific system structure. Generally, the proposed sensor categorization approach provides a more effective solution when a CPS consists of a few types of sensors.

\section{Experimental Results}

\subsection{Search Space Reduction}

To show the combinatorial complexity reduction by using the equivalent class approach, we perform a comparison on different equivalence partitions by setting $p=6$ and $s=2$. Then for the sensor set $\mathbb{S}=\{\mathbf{S}_i,i=1,\cdots,6\}$, we define
$$
\begin{aligned}
&\Gamma_{12}=\{\mathbf{S}_1,\mathbf{S}_2\},~\Gamma_{13}=\{\mathbf{S}_1,\mathbf{S}_3\},~\Gamma_{14}=\{\mathbf{S}_1,\mathbf{S}_4\},\\
&\Gamma_{15}=\{\mathbf{S}_1,\mathbf{S}_5\},~\Gamma_{16}=\{\mathbf{S}_1,\mathbf{S}_6\},~\Gamma_{23}=\{\mathbf{S}_2,\mathbf{S}_3\},\\
&\Gamma_{24}=\{\mathbf{S}_2,\mathbf{S}_4\},~\Gamma_{25}=\{\mathbf{S}_2,\mathbf{S}_5\},~\Gamma_{26}=\{\mathbf{S}_2,\mathbf{S}_6\},\\
&\Gamma_{34}=\{\mathbf{S}_3,\mathbf{S}_4\},~\Gamma_{35}=\{\mathbf{S}_3,\mathbf{S}_5\},~\Gamma_{36}=\{\mathbf{S}_3,\mathbf{S}_6\},\\
&\Gamma_{45}=\{\mathbf{S}_4,\mathbf{S}_5\},~\Gamma_{46}=\{\mathbf{S}_4,\mathbf{S}_6\},~\Gamma_{56}=\{\mathbf{S}_5,\mathbf{S}_6\}.
\end{aligned}
$$
Then $\Sigma=\{\Gamma_{ij},1\le i<j\le6\}$ and the size of search space is 15. TABLE I shows that different search sets for different equivalence partitions under $\mho_{\mathcal{T}}^\diamond=\emptyset$. Clearly different attack models cause different search space. By computation, the average size of search space for different partitioning cases in TABLE I is 5. Clearly the combinatorial search space has been narrowed by 66.7\%. Also, it can be summarized from TABLE I that in general, the search space can be reduced as the rank of the Mean-Agreeable sensor type increases.

\begin{figure*}
\begin{center}
{\bf TABLE~I} ~Sizes of search space for different cases
\label{tab:2} \vskip 7pt
\newcommand{\rb}[1]{\raisebox{1.5ex}[-1pt]{#1}}
\begin{tabular}{|c c c c|}\hline
\multicolumn{1}{|c|}{RE-partition $\mho_{\mathcal{T}}$}& \multicolumn{1}{|c|}{Mean-Agreement $\mho_{\mathcal{T}}^+$} & \multicolumn{1}{|c|}{ Search space $\Sigma_{\mathcal{T}}$ ($|\Sigma_{\mathcal{T}}|$)}\\
\hline \multicolumn{1}{|c|}{\{$\mathbf{S}_1,\mathbf{S}_2,\mathbf{S}_3\}$,\{$\mathbf{S}_4,\mathbf{S}_5,\mathbf{S}_6$\}}& \multicolumn{1}{|c|}{\{$\mathbf{S}_1,\mathbf{S}_2,\mathbf{S}_3\}$} & \multicolumn{1}{|c|}{$\Gamma_{45},\Gamma_{45},\Gamma_{56} (3)$} \\
\hline \multicolumn{1}{|c|}{\{$\mathbf{S}_1,\mathbf{S}_2,\mathbf{S}_3\}$,\{$\mathbf{S}_4,\mathbf{S}_5,\mathbf{S}_6$\}}& \multicolumn{1}{|c|}{$\emptyset$} & \multicolumn{1}{|c|}{$\Gamma_{ij},i=1,2,3,j=4,5,6$(9)}\\
\hline
\multicolumn{1}{|c|}{$\Gamma_{12},\Gamma_{34},\Gamma_{56}$}& \multicolumn{1}{|c|}{$\Gamma_{12},\Gamma_{34},\Gamma_{56}$} &\multicolumn{1}{|c|}{$\Gamma_{12},\Gamma_{34},\Gamma_{56}$ (3)}\\
\hline
\multicolumn{1}{|c|}{$\Gamma_{12},\Gamma_{34},\Gamma_{56}$}& \multicolumn{1}{|c|}{$\Gamma_{12},\Gamma_{34}$} &\multicolumn{1}{|c|}{$\Gamma_{56}$ (1)}\\
\hline
\multicolumn{1}{|c|}{$\Gamma_{12},\Gamma_{34},\Gamma_{56}$}& \multicolumn{1}{|c|}{$\Gamma_{12}$} &\multicolumn{1}{|c|}{$\Gamma_{ij},i=3,4,j=5,6$ (4)}\\
\hline
\multicolumn{1}{|c|}{\{$\mathbf{S}_1\},\{\mathbf{S}_2\},\{\mathbf{S}_3,\mathbf{S}_4,\mathbf{S}_5,\mathbf{S}_6$\}}& \multicolumn{1}{|c|}{$\{\mathbf{S}_3,\mathbf{S}_4,\mathbf{S}_5,\mathbf{S}_6\}$} & \multicolumn{1}{|c|}{$\Gamma_{12}$ (1)}\\
\hline
\multicolumn{1}{|c|}{\{$\mathbf{S}_1\},\{\mathbf{S}_2\},\{\mathbf{S}_3,\mathbf{S}_4,\mathbf{S}_5,\mathbf{S}_6$\}}& \multicolumn{1}{|c|}{$\emptyset$} & \multicolumn{1}{|c|}{$\Sigma\setminus\{\Gamma_{12}\}$ (14)} \\
\hline
\multicolumn{1}{|c|}{\{$\mathbf{S}_1\},\{\mathbf{S}_2\},\{\mathbf{S}_3\},\{\mathbf{S}_4\},\Gamma_{56}$}& \multicolumn{1}{|c|}{$\Gamma_{56}$} & \multicolumn{1}{|c|}{$\Gamma_{56},\Gamma_{ij},1\le i<j\le4$ (5)} \\
\hline
\multicolumn{1}{|c|}{\{$\mathbf{S}_1\},\{\mathbf{S}_2\},\{\mathbf{S}_3\},\{\mathbf{S}_4\},\Gamma_{56}$}& \multicolumn{1}{|c|}{$\emptyset$} & \multicolumn{1}{|c|}{$\Gamma_{ij},i=1,2,3,4,j=5,6$ (8)} \\
\hline
\multicolumn{1}{|c|}{\{$\mathbf{S}_1\},\{\mathbf{S}_2\},\Gamma_{34},\Gamma_{56}$}& \multicolumn{1}{|c|}{$\Gamma_{34},\Gamma_{56}$} & \multicolumn{1}{|c|}{$\Gamma_{12},\Gamma_{34},\Gamma_{56}$ (3)}\\
\hline
\multicolumn{1}{|c|}{\{$\mathbf{S}_1\},\{\mathbf{S}_2\},\Gamma_{34},\Gamma_{56}$}& \multicolumn{1}{|c|}{$\Gamma_{34}$} & \multicolumn{1}{|c|}{$\Gamma_{56},\Gamma_{ij},i=1,2,j=5,6$ (5)}\\
\hline
\multicolumn{1}{|c|}{\{$\mathbf{S}_1\},\{\mathbf{S}_2\},\Gamma_{34},\Gamma_{56}$}& \multicolumn{1}{|c|}{$\emptyset$} & \multicolumn{1}{|c|}{$\Gamma_{ij},i=3,4,j=5,6$ (4)} \\
\hline
\end{tabular}\\
\end{center}
\end{figure*}

In the other case, if $\mho_{\mathcal{T}}^\diamond\ne\emptyset$, i.e., there exists an equivalent class $[\mathbf{S}_i]_{\mathcal{T}}$ such that $|[\mathbf{S}_i]_{\mathcal{T}}|\ge 5$, then we can select $\Sigma_{\mathcal{T}}=\mathbb{S}\setminus[\mathbf{S}_i]_{\mathcal{T}}^\oplus$ under the conditions of $4$-sparse observability and $|\mathbb{S}\setminus[\mathbf{S}_i]_{\mathcal{T}}^\oplus|\ge 3$. Resultantly, the search space size is only one.

It is known that the proposed SSE method and the existing set cover approach \cite{AY2017} have a similar structure:  pre-determining the sensor set, as a way of reducing computational complexity at runtime. To illustrate the advantage of the proposed sensor categorization technique over the set cover technique, we consider the cases of $p=10,12,14,16,18,20$ and $s=2$, and assume that every two sensors belong to the same sensor type. According to Theorem 1, there are the following three cases:
\begin{description}
  \item[Case 1:] There are two sensor types which are not Mean-Agreement. Then, the size of search space is 4;
  \item[Case 2:] There are only one sensor type which is not Mean-Agreement. Then, the size of search space is 1;
  \item[Case 3:] All sensor types are not Mean-Agreement. Then, the size of search space is $p/2$.
\end{description}
Summarizing the above three cases, the average size of search space is $\lceil(10+p)/6\rceil$.

The average sizes of search space for different SSE methods are shown in the following table.

\begin{center}\small
{\bf TABLE~II} ~Sizes of search space for different SSE methods
\label{tab:2} \vskip -5pt
\newcommand{\rb}[1]{\raisebox{1.5ex}[-1pt]{#1}}
\begin{tabular}{|c c c c c c c|}\hline
\multicolumn{1}{|c|}{Number of sensors $p$} & \multicolumn{1}{|c|}{10} & \multicolumn{1}{|c|}{12}& \multicolumn{1}{|c|}{14}
& \multicolumn{1}{|c|}{16}& \multicolumn{1}{|c|}{18}& \multicolumn{1}{|c|}{20}\\
\hline \multicolumn{1}{|c|}{Exhaustive search} & \multicolumn{1}{|c|}{45} & \multicolumn{1}{|c|}{66} & \multicolumn{1}{|c|}{91}& \multicolumn{1}{|c|}{120} & \multicolumn{1}{|c|}{153}& \multicolumn{1}{|c|}{190}\\
\hline \multicolumn{1}{|c|}{Set cover approach \cite{AY2017}} & \multicolumn{1}{|c|}{9} & \multicolumn{1}{|c|}{10}& \multicolumn{1}{|c|}{12}
& \multicolumn{1}{|c|}{14}& \multicolumn{1}{|c|}{16}& \multicolumn{1}{|c|}{20}\\
\hline \multicolumn{1}{|c|}{The proposed approach} & \multicolumn{1}{|c|}{4} & \multicolumn{1}{|c|}{4} & \multicolumn{1}{|c|}{4}& \multicolumn{1}{|c|}{5} & \multicolumn{1}{|c|}{5}& \multicolumn{1}{|c|}{5}\\
\hline
\end{tabular}\\
\end{center}

It can be seen that, these two approaches both narrow the combinatorial search space, and the proposed senor categorization technique has smaller search space than the existing set cover approach \cite{AY2017}.

\subsection{Performance Verification}

In order to verify the computation performance of the proposed scheme, a three-inertia system consisting of three rigid inertias coupled by two flexible shafts is simulated.
Borrowing from  \cite{CL2015}, we describe the dynamics by
\begin{equation}
\begin{aligned}
\dot x=&A_c x+B_cT_m+B_c^dT_L\\
y=&C_cx+n
\end{aligned}
\end{equation}
with
$$
A_c=\left[\begin{matrix}
0 & 1 & 0 & 0 & 0 & 0\\
-\frac{K_1}{J_1} & -\frac{B_1}{J_1} & \frac{K_1}{J_1} & 0 & 0 & 0\\
0 & 0 & 0 & 1 & 0 & 0\\
\frac{K_1}{J_2} & 0 & -\frac{K_1+K_2}{J_2} & -\frac{B_2}{J_2} & \frac{K_2}{J_2} & 0\\
0 & 0 & 0 & 0 & 0 & 1\\
0 & 0 & \frac{K_2}{J_3} & 0 & -\frac{K_2}{J_3} & -\frac{B_3}{J_3}
\end{matrix}\right],
$$
and $B_c=\left[0~1/J_1~0~0~0~0\right]^T, B_c^d=\left[0~0~0~0~0~1/J_3\right]^T$, where\\
$~~J_1:$ inertia of drive motor\\
$~~J_2:$ inertia of middle body\\
$~~J_3:$ inertia of load\\
$~~\theta_1,\theta_2,\theta_3:$ absolute angular positions of three inertias\\
$~~\omega_1,\omega_2, \omega_3:$ speeds of three inertias\\
$~~K_1,K_2:$ torsional stiffness of two shafts\\
$~~B_1,B_2,B_3:$ mechanical damping of three inertias\\
$~~T_m:$ motor drive torque\\
$~~T_L$: load disturbance torque\\
and the state vector is $x=[\theta_1~\omega_1~\theta_2~\omega_2~\theta_3~\omega_3]^T$ and the output measurements are the absolute angular positions of three inertias $\theta_1,\theta_2,\theta_3$ and three extra measurements which are the relative angles $\theta_1-\theta_2,\theta_1-\theta_3$ and $\theta_2-\theta_3$ between any two inertias, $n$ represents the measurement noise bounded by $\|n\|\le0.001$.
Then the output matrix is
$$
C_c=\left[\begin{matrix}
1 & 0 & 0 & 0 & 0 & 0\\
0 & 0 & 1 & 0 & 0 & 0\\
0 & 0 & 0 & 0 & 1 & 0\\
1 & 0 & -1 & 0 & 0 & 0\\
1 & 0 & 0 & 0 & -1 & 0\\
0 & 0 & 1 & 0 & -1 & 0\\
\end{matrix}\right].
$$
Assume $\|T_L\|\le0.01$. In the simulation, $J_1=J_2=J_3=0.01$kg$\cdot$m$^2$, $B_1=B_2=B_3=0.005$N/(rad/s) and $K_1=K_2=1.4$N/rad.
To obtain the discrete-time model of the three-inertia system, we discretize
the continuous-time model by taking periodic sampling with
sampling period $h = 0.1$s. In the simulation, choose $x(0)=[-0.2,0.1,0,0.3,0.1,0.2]^T$. Then $\Sigma=\{\Gamma_{ij},1\le i<j\le6\}$ which is defined in Section IV-A and the size of search space is 15. The control input is set as $T_m=K\hat x+\sin(0.2t)$ where $K=[0.7732~-0.0718~-0.8379~-0.0351~-0.0137~-0.0213]$.

We perform the DE-Partitioning by following Algorithm 3. First, the ranks of all $\mathcal{O}_i$, $i=1,\cdots,6$ are obtained as
$$
\begin{aligned}
&\mathrm{rank}(\mathcal{O}_1)=\mathrm{rank}(\mathcal{O}_3)=6,\\
&\mathrm{rank}(\mathcal{O}_4)=\mathrm{rank}(\mathcal{O}_6)=4,\\
&\mathrm{rank}(\mathcal{O}_2)=5,\mathrm{rank}(\mathcal{O}_5)=2.\\
\end{aligned}
$$
By solving LMI (6) with $\varepsilon=10^{-4}$, we obtain that $\mathbf{S}_1\overset{\mathcal{T}}{\sim}\mathbf{S}_3$ with $T_{13}=\mathcal{O}_1\mathcal{O}_3^{-1}$ and $\mathbf{S}_4\overset{\mathcal{T}}{\sim}\mathbf{S}_6$ with $T_{46}=$
$$
\left[\begin{matrix}
 -0.0394 &   0.1480 &  -0.0399 &  -0.5771 &   0.1551 &   0.5628\\
 0.4629  &  0.0480  &  0.2747  & -0.0309  & -0.7617  &  0.0751\\
 0.0071  &  0.4763  & -0.0082  &  0.2813  & -0.1140  & -0.7602\\
 -0.7148 &  -0.1080 &   0.1856 &  -0.0115 &   0.4349 &   0.0139\\
 0.0708  & -0.7146  & -0.0254  &  0.1800  &  0.0476  &  0.4197\\
 0.4550  &  0.1324  & -0.4734  & -0.0296  &  0.1598  & -0.0365
\end{matrix}\right].
$$
Hence, we obtain the equivalence partition of set $\mathbb{S}$
$$
\mho_{\mathcal{T}}=\left\{\{\mathbf{S}_2\},\{\mathbf{S}_5\},\{\mathbf{S}_1,\mathbf{S}_3\},\{\mathbf{S}_4,\mathbf{S}_6\}\right\}.
$$
TABLE II gives a detailed analysis on the search space under different cases of Mean-Agreeable sensor types.

\begin{center}
{\bf TABLE~II} ~Search space for Mean-Agreeable types
\label{tab:2} \vskip -5pt
\newcommand{\rb}[1]{\raisebox{1.5ex}[-1pt]{#1}}
\begin{tabular}{|c c c c|}\hline
\multicolumn{1}{|c|}{Agreeable type $(\mho_{\mathcal{T}}^+)$} & \multicolumn{1}{|c|}{ Search space ($\Sigma_{\mathcal{T}}$)} & \multicolumn{1}{|c|}{\it $|\Sigma_{\mathcal{T}}|$}\\
\hline \multicolumn{1}{|c|}{$\{\mathbf{S}_1,\mathbf{S}_3\},\{\mathbf{S}_4,\mathbf{S}_6$\}} & \multicolumn{1}{|c|}{$\Gamma_{13},\Gamma_{46},\Gamma_{25}$} & \multicolumn{1}{|c|}{3}\\
\hline \multicolumn{1}{|c|}{$\{\mathbf{S}_1,\mathbf{S}_3\}$} & \multicolumn{1}{|c|}{$\Gamma_{24},\Gamma_{26},\Gamma_{45},\Gamma_{46},\Gamma_{56}$} & \multicolumn{1}{|c|}{5}\\
\hline \multicolumn{1}{|c|}{$\{\mathbf{S}_4,\mathbf{S}_6\}$} & \multicolumn{1}{|c|}{$\Gamma_{13},\Gamma_{12},\Gamma_{15},\Gamma_{23},\Gamma_{35}$} & \multicolumn{1}{|c|}{5}\\
\hline \multicolumn{1}{|c|}{$\emptyset$} & \multicolumn{1}{|c|}{$\Gamma_{ij},i=1,3,j=4,6$} & \multicolumn{1}{|c|}{4}\\
\hline
\end{tabular}\\
\end{center}

We consider the following 4 attack models:
\begin{description}
  \item[Case 1:] Sensors $\mathbf{S}_5$ and $\mathbf{S}_6$ are attacked and $a_5(t)$ and $a_6(t)$ are randomly generated from $[0,2]$.
  \item[Case 2:] Sensors $\mathbf{S}_2$ and $\mathbf{S}_5$ are attacked and $a_2(t)$ and $a_5(t)$ are randomly generated from $[-5,5]$.
  \item[Case 3:] Sensors $\mathbf{S}_3$ and $\mathbf{S}_6$ are attacked and $a_3(t)$ and $a_6(t)$ are randomly generated from $[0,2]$.
  \item[Case 4:] Sensors $\mathbf{S}_4$ and $\mathbf{S}_6$ are attacked and $a_4(t)$ and $a_6(t)$ are randomly generated from $[0,1]$.
\end{description}

Now we perform the EX-SEARCH($\Sigma_{\mathcal{T}},\eta\Delta_s$) in Algorithm 1. To reduce the computation complexity, we adopt the Mean-Agreement instead of Median-Agreement (refer to Remark 2). Figs. 1-4 show the estimation performance ($\theta_3$ and its estimate $\hat\theta_3$) and the sizes of search space under four cases. In Case 1, the attack $a_6$ causes that sensor type $\{\mathbf{S}_4,\mathbf{S}_6\}$ is not Mean-Agreeable and only $\{\mathbf{S}_1,\mathbf{S}_3\}$ is Mean-Agreeable most of time. Referring to TABLE II, the size of search space is 5. However, sometimes the attack signal $a_6$ is small (undetectable) such that $\{\mathbf{S}_4,\mathbf{S}_6\}$ is still Median-Agreeable, and the size of search space is 3. In Case 2, sensor types $\{\mathbf{S}_4,\mathbf{S}_6\}$ and  $\{\mathbf{S}_1,\mathbf{S}_3\}$ both are Median-Agreeable all the time, and the size of search space is 3. In Case 3, the attacks $a_3$ and $a_6$ cause that either $\{\mathbf{S}_4,\mathbf{S}_6\}$ or  $\{\mathbf{S}_1,\mathbf{S}_3\}$ is not Mean-Agreeable most of time, and the size of search space is 4. However, at some time, the attack signal is small such that one of two sensor types is still Mean-Agreeable. Under this case, the size of search space is 5. Note that the probability that both sensor types are simultaneously Mean-Agreeable under the attacks (the corresponding size of search space is 3) is very small  and this case did not happen in the simulation. In Case 4, two attack signals simultaneously occur on a sensor type $\{\mathbf{S}_4,\mathbf{S}_6\}$. Similar to the previous analysis, the size of search space is switched between 3 and 5. Additionally, from Figs. 1, 3 and 4, the undetectable attack may cause an increase of estimate error (referring to Theorem 2).

\begin{figure}
  \centering
  \includegraphics[width=8.5cm,height=5.3cm]{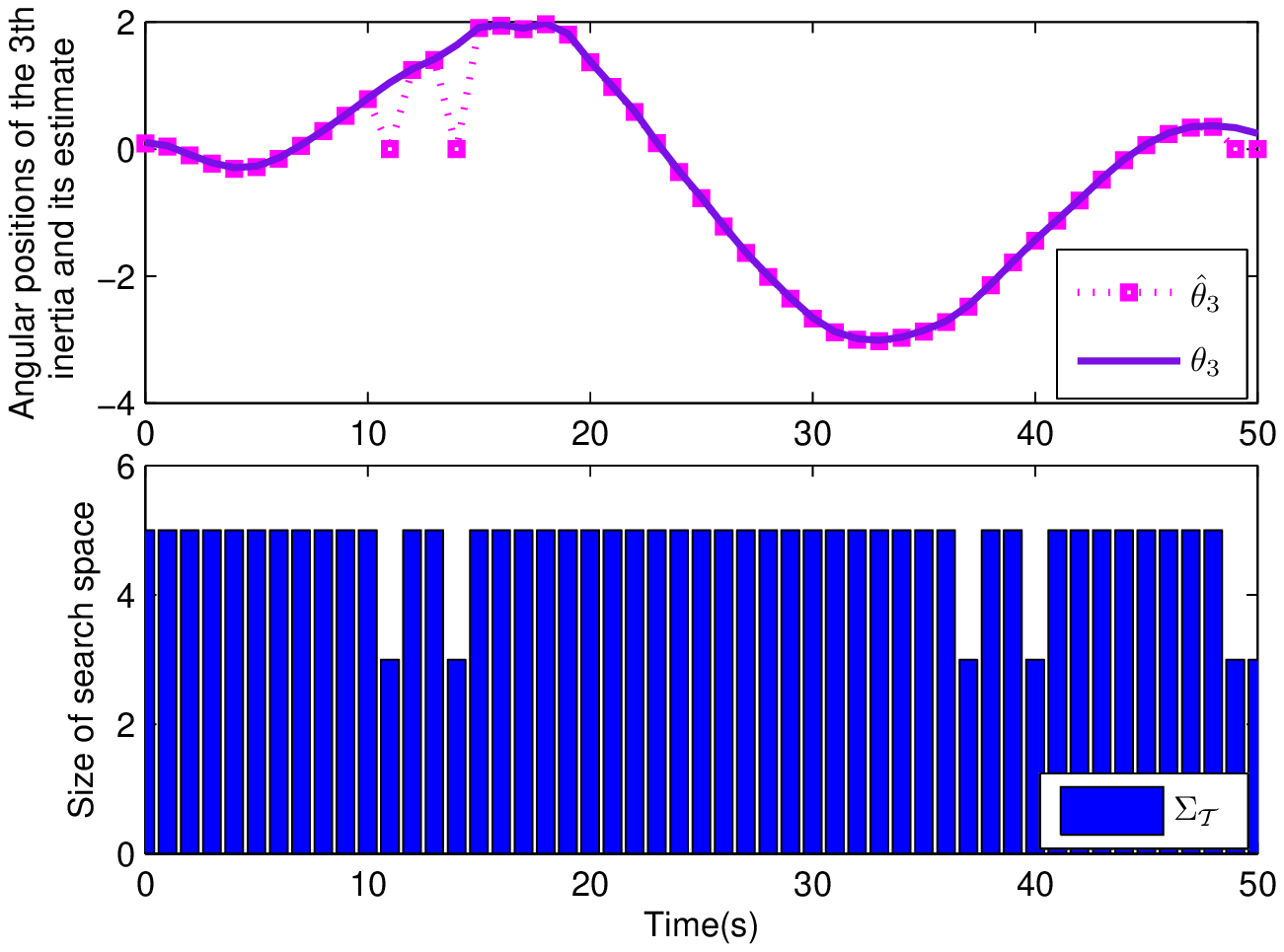}\\
  \caption{Simulation results for Case 1.}
      \centering
  \includegraphics[width=8.5cm,height=5.3cm]{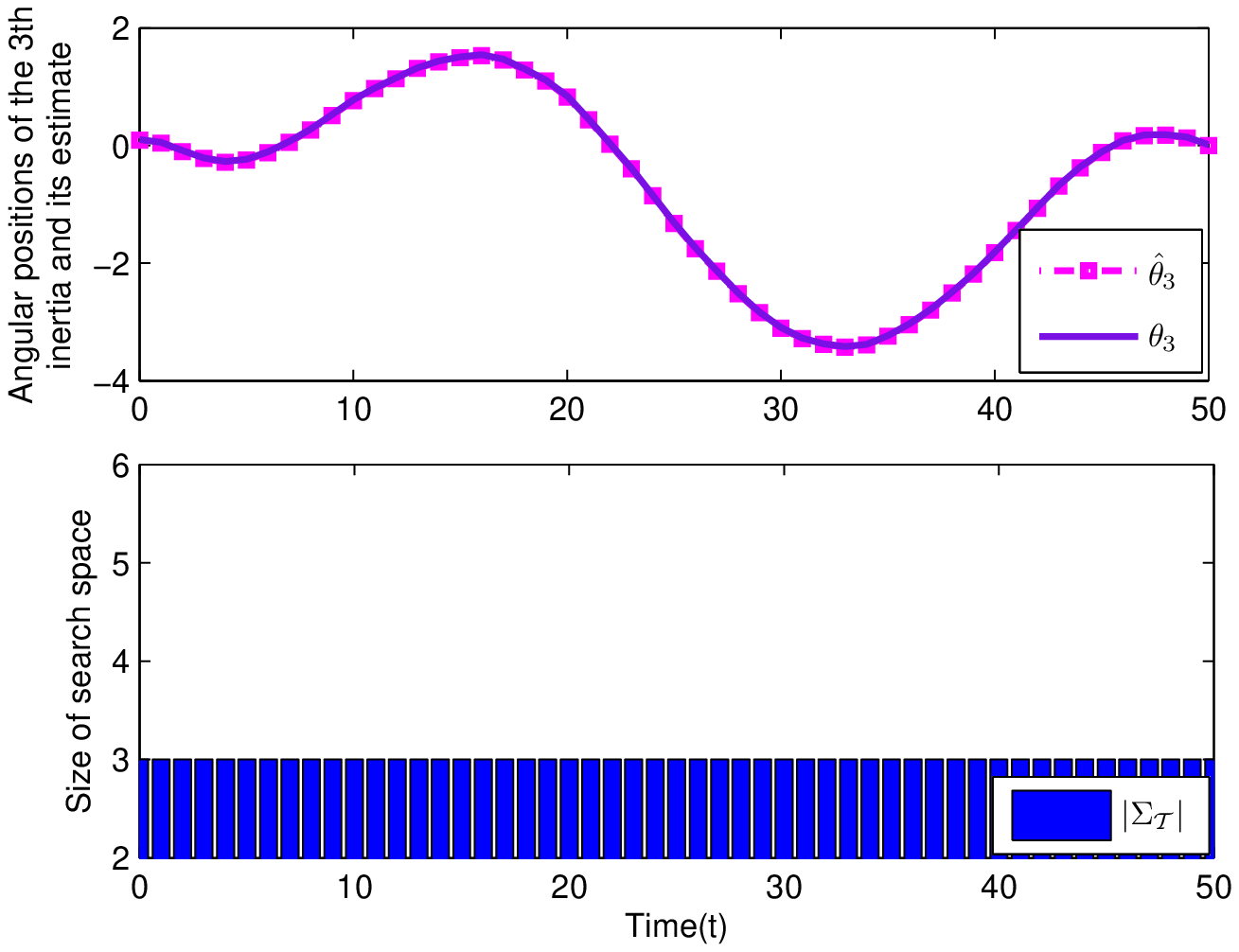}\\
  \caption{Simulation results for Case 2.}
  \centering
  \includegraphics[width=8.5cm,height=5.3cm]{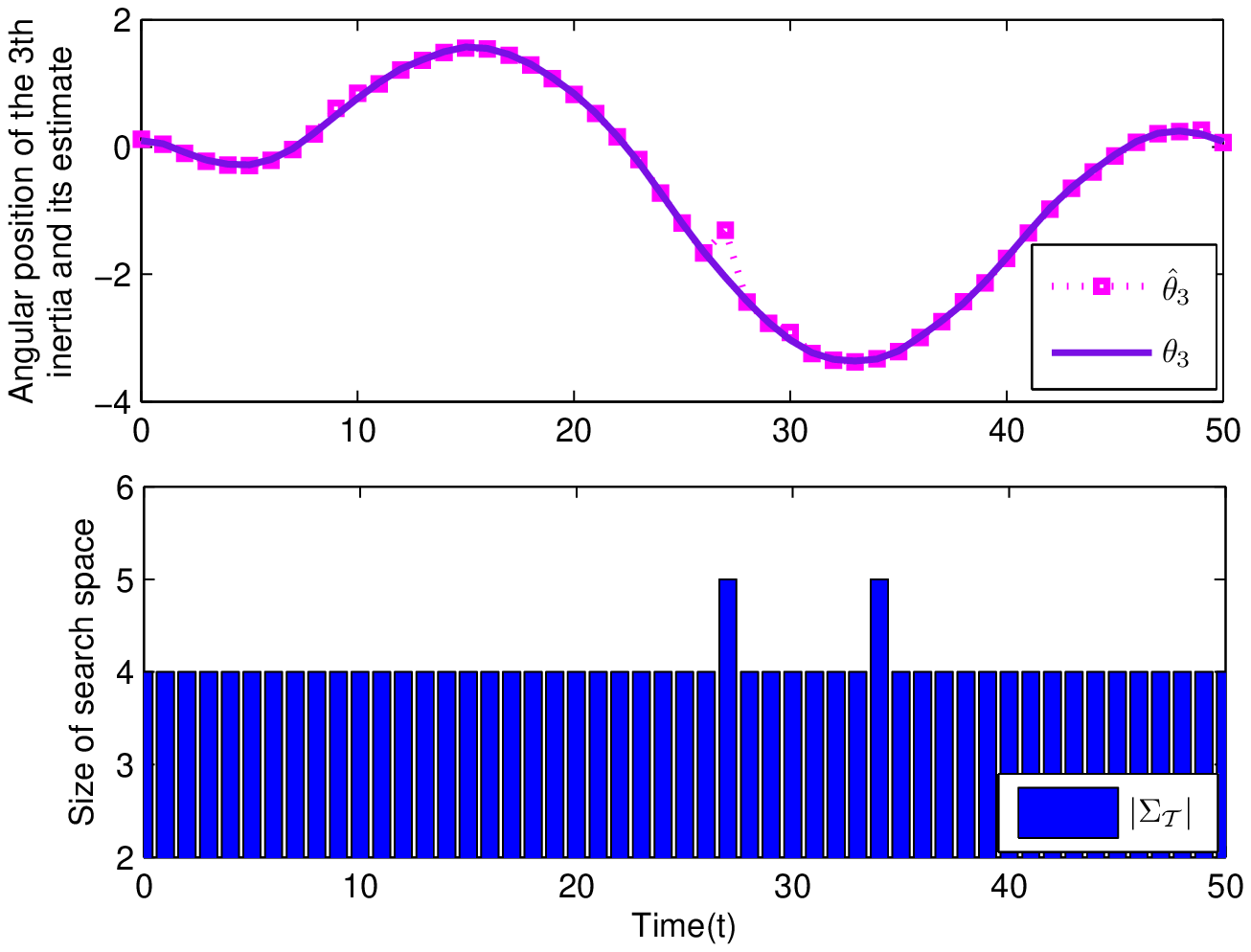}\\
  \caption{Simulation results for Case 3.}
\end{figure}

\begin{figure}
      \centering
  \includegraphics[width=8.5cm,height=5.3cm]{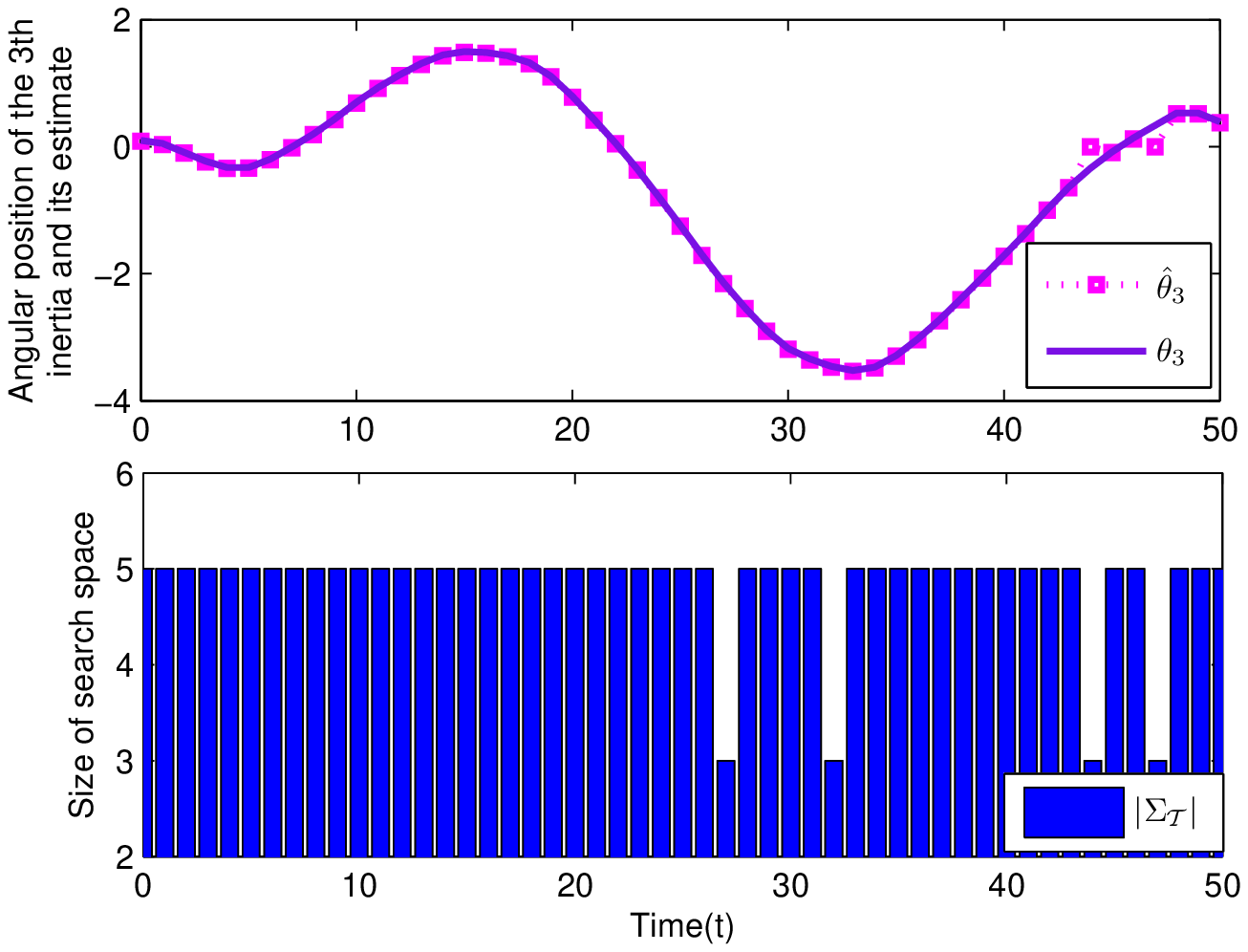}\\
  \caption{Simulation results for Case 4.}
  \centering
  \includegraphics[width=8.5cm,height=5.3cm]{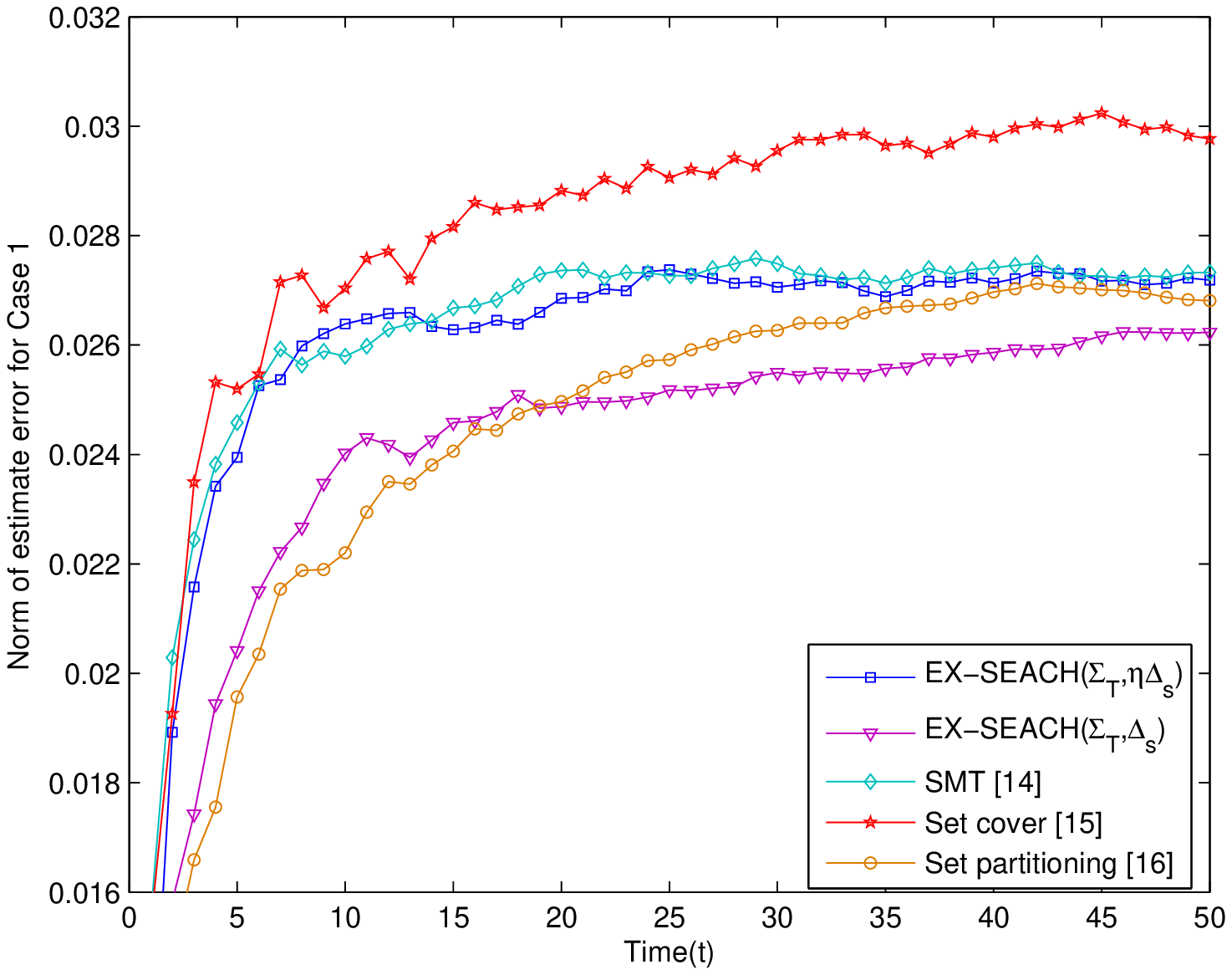}\\
  \caption{Norm of estimate error under Case 1 for different algorithms}
\end{figure}

As reported in the introduction, we can divide these existing SSE algorithms into two categories according to the way of reducing computation complexity: i) relaxation \cite{YH2015,CS2017,HF2014,MP2017,YS2016} and ii) narrowing search space \cite{YS2017,AY2017,LA2018}. The proposed algorithm belongs to the second category, for fairness, we only compare the proposed algorithm (EX-SEARCH($\Sigma_{\mathcal{T}},\eta\Delta_s$) with the second category (EX-SEARCH($\Sigma,\Delta_s$), SMT approach \cite{YS2017}\footnote[1]{Here we have modified the operation $x:=\arg\min_{x_\in \mathbb{R}^n}\|Y_\mathcal{I}-\mathcal{O}_\mathcal{I}x\|_2^2$ as $x:=\mathcal{O}_\mathcal{I}^+Y_\mathcal{I}$ in the Algorithm 2 of SMT approach \cite{YS2017} such that the continuous and infinite search space is transformed into discrete and finite one.}, set cover approach \cite{AY2017}, constrained set partitioning approach\cite{LA2018}) {\it w.r.t.} computational and estimation performances. The programs ran in MATLAB on a desktop equipped with an Intel Core i7-6700 processor operating at 3.4 GHz and 4 GB of RAM.

We define the {\it execution time} as the time needed by an algorithm to compute an estimate at a time instant. The execution time\footnote[2]{The execution time and the norms of errors are taken as the average values
of them for 100 tests.} of these seven algorithms under different cases is given in TABLE III. It can be seen that the proposed algorithm takes less execution time than that of the existing methods. Note that although the set cover approach proposed in \cite{AY2017} only needs to test 3 set candidates $\{\mathbf{S}_1,\mathbf{S}_2,\mathbf{S}_3,\mathbf{S}_4\}$, $\{\mathbf{S}_1,\mathbf{S}_2,\mathbf{S}_5,\mathbf{S}_6\}$ and $\{\mathbf{S}_3,\mathbf{S}_4,\mathbf{S}_5,\mathbf{S}_6\}$, the used sparsity projector $\Pi$ (see Definition 3 of \cite{AY2017}) still requires a relatively long execution time.

\begin{center}
{\bf TABLE~III} ~Execution time [ms] under five cases
\label{tab:2} \vskip -5pt
\newcommand{\rb}[1]{\raisebox{1.5ex}[-1pt]{#1}}
\begin{tabular}{|c c c c c c|}\hline
\multicolumn{1}{|c|}{No.} & \multicolumn{1}{|c|}{Case 1} & \multicolumn{1}{|c|}{Case 2} & \multicolumn{1}{|c|}{Case 3} & \multicolumn{1}{|c|}{Case 4}\\
\hline \multicolumn{1}{|c|}{1} & \multicolumn{1}{|c|}{4.99} & \multicolumn{1}{|c|}{4.51} & \multicolumn{1}{|c|}{4.94} & \multicolumn{1}{|c|}{4.83}\\
\hline \multicolumn{1}{|c|}{2} & \multicolumn{1}{|c|}{15.51} & \multicolumn{1}{|c|}{8.33} & \multicolumn{1}{|c|}{13.40} & \multicolumn{1}{|c|}{14.27}\\
\hline \multicolumn{1}{|c|}{3} & \multicolumn{1}{|c|}{12.71} & \multicolumn{1}{|c|}{5.47} & \multicolumn{1}{|c|}{9.44} & \multicolumn{1}{|c|}{10.43}\\
\hline \multicolumn{1}{|c|}{4} & \multicolumn{1}{|c|}{5.61} & \multicolumn{1}{|c|}{6.30} & \multicolumn{1}{|c|}{6.59} & \multicolumn{1}{|c|}{6.77}\\
\hline \multicolumn{1}{|c|}{5} & \multicolumn{1}{|c|}{12.57} & \multicolumn{1}{|c|}{7.85} & \multicolumn{1}{|c|}{5.66} & \multicolumn{1}{|c|}{10.61}\\
\hline
\end{tabular}\\
\end{center}
where No. 1: EX-SEARCH($\Sigma_{\mathcal{T}},\eta\Delta_s$); No. 2: EX-SEARCH($\Sigma,\Delta_s$);  No. 3: SMT  \cite{YS2017}; No. 4: Set cover \cite{AY2017}; No. 5: Set partitioning \cite{LA2018}.

For illustrating the estimation performance, the norms of estimate errors $\|x(t)-\hat x(t)\|$ of these five SSE algorithms only under Case 1 are shown in Fig. 5. It can be seen that the set cover based approach \cite{AY2017} has a lower estimation accuracy than the others (Reduced by 11.1\% compared with the proposed sensor categorization approach) due to the information loss caused by the remove of additional two attack-free sensor measurements.

The above comparison results confirm high computational efficiency and estimation accuracy of the proposed SSE algorithm.

It is worth noting that, the four algorithms in \cite{YS2017,AY2017,LA2018} and this paper are all used for reducing the search candidates in different fashions and a more effective approach, if possible, is the combination of them. In this example, the proposed sensor categorization approach provides a lower-complexity solution than the others since some sensors in system (9) have the analytic observation equivalence. Also, this property is also satisfied by many other practical systems.

\section{Conclusion}

In this paper, we have proposed a fast secure estimation algorithm based on an equivalent class approach. A concept of analytic sensor types has been introduced to describe the measurement equivalence of sensors. By verifying the measurement data of the sensor types, the combinatorial search space has been narrowed. The high speed performance of the proposed algorithm has been illustrated by comparative simulation results.
Motivated by \cite{QH2018,SW2012,JK2019}, future work focuses on extending the proposed SSE method to the nonlinear systems. The identification of sensor faults and attacks and designs of different tolerance strategies are another interesting issues.



\begin{thebibliography}{99}


\bibitem{SS2012}
S. Sridhar, A. Hahn, and M. Govindarasu, ``Cyber physical system security for the electric
power grid,'' {\it Proceedings of the IEEE}, vol. 100, no. 1, pp.210--224, 2012.

\bibitem{CL2015}
C. Lee, H. Shim, and Y. Eun, ``On redundant observability: from security index to attack detection and resilient state estimation,'' {\it IEEE Trans. Autom. Control}, vol. 64, no. 2, pp. 775--782.


\bibitem{FP2013}
F. Pasqualetti, F. Dorfler, and F. Bullo, ``Attack detection and identification in cyber-physical systems,'' {\it IEEE Trans. Autom. Control}, vol. 58, no. 11, pp. 2715--2729, 2013


\bibitem{MS2015}
M.S. Chong, M. Wakaiki and J.P. Hespanha, ``Observability of linear systems under adversarial attacks,'' {\it in Proc. American Control Conf. Palmer House Hilton Chicago, USA}, 2015, pp. 2439--2444.

\bibitem{SM2014}
S. Mishra, N. Karamchandani, P.  Tabuada, and  S. Diggavi, ``Secure state
estimation and control using multiple (insecure) observers,'' {\it  in Proc. 53rd IEEE Conf. Decision and Control}, Los Angeles, California, USA, 2014,  pp. 1620-1625.

\bibitem{LW2017}
L. An and G.-H. Yang, ``Secure state estimation against sparse sensor attacks with adaptive switching mechanism,'' {\it IEEE Trans. Autom. Control}, vol. 63, no. 8, pp. 2596--2603, 2018.


\bibitem{ZY2018}
Z. Guo, D. Shi, D.E. Quevedo, L. Shi, ``Secure state estimation against integrity attacks: a
gaussian mixture model approach,'' {\it IEEE Trans. Signal Process.}, vol. 67, no. 1, pp. 194--207.

\bibitem{YH2015}
Y.H. Chang, Q. Hu, and C.J. Tomlin, ``Secure estimation based
Kalman filter for cyber-physical systems against adversarial attacks,''
{\it Automatica}, vol. 95, pp. 399--412, 2018.

\bibitem{CS2017}
C. Liu, J. Wu, C. Long, and Y. Wang, ``Dynamic state recovery for cyber-physical systems under switching location attacks,'' {\it IEEE Trans. Control Netw. Syst.,} vol. 4, no. 1, pp. 14--22, 2017.

\bibitem{HF2014}
H. Fawzi, P. Tabuada, and S. Diggavi, ``Secure estimation and control for cyber-physical systems under adversarial attacks,'' {\it IEEE Trans. Autom. Control}, vol. 59, no. 6, pp. 1454--1467, 2014.

\bibitem{MP2017}
M. Pajic, I. Lee, and G.J. Pappas, ``Attack-resilient state estimation for noisy
dynamical systems,'' {\it IEEE Trans. Control Netw. Syst.,} vol. 4, no. 1, pp. 82--92, 2017.

\bibitem{YS2016}
Y. Shoukry and P. Tabuada, ``Event-triggered state observers for sparse sensor noise/attacks,'' {\it IEEE Trans. Autom. Control}, vol. 61, no. 8, 2079--2091, 2016.

\bibitem{YS2017}
Y. Shoukry, P. Nuzzo, A. Puggelli, A.L. Sangiovanni-Vincentelli, S.A. Seshia, and P. Tabuada, ``Secure state estimation for cyber physical systems under sensor
attacks: a satisfiability modulo theory approach,'' {\it IEEE Trans. Autom. Control}, vol. 62, no. 10, pp. 4917--4932, 2017.

\bibitem{AY2017}
A-Y. Lu and G-H. Yang,  ``Secure switched observers for cyber-physical systems under sparse sensor attacks: a set cover approach,'' {\it IEEE Trans. Autom. Control}, vol. 64, no. 9, pp. 3949--3955, 2019.

\bibitem{LA2018}
L. An and G-H. Yang,  ``State estimation under sparse sensor attacks: a constrained set partitioning approach,'' {\it IEEE Trans. Autom. Control}, vol. 64, no. 9, pp. 3861--3868.



\bibitem{SA2011}
S. Arai, Y. Iwatani, and K. Hashimoto, ``Fast sensor scheduling for spatially distributed sensors,'' {\it IEEE Trans. Autom. Control}, vol. 56, no. 8, pp. 1900--1905, 2011.

\bibitem{IS2005}
I. Stojmenovic, Sensor Networks. New York: Wiley InterScience,
2005.

\bibitem{KS2010}
K. Segl, L. Guanter, H. Kaufmann, J. Schubert, S. Kaiser, B. Sang, and S. Hofer, ``Simulation of spatial sensor characteristics in the context of the EnMAP hyperspectral mission,'' {\it IEEE Trans. Geos. Remo. Sens.}, vol. 48, no. 7, pp. 3046--3054, 2010.

\bibitem{HM1994}
M. Hazewinkel, ed. ``Equivalence relation", {\it Encyclopedia of Mathematics}, Springer Science+Business Media B.V./ Kluwer Academic Publishers, ISBN 978-1-55608-010-4, 1994.

\bibitem{BL2003}
B.L. Stevens and F.L. Lewis, {\it Aircraft control and simulation}. John wiley \& sons,
Inc. 2003.

\bibitem{HC1970}
C. Hanke and D. Nordwall, ``The simulation of a jumbo jet transport
aircraft'', {\it Modelling data NASA and the Boeing Company}, vol. II.
Technical report CR-114494/D6-30643-VOL2, 1970.

\bibitem{W}
``Median of medians,'' {\it Wikipedia}, Available: https://en. wikipedia.org/wiki/Median\_of\_medians

\bibitem{QH2018}
Q. Hu, D. Fooladivanda , Y.H. Chang , and C.J. Tomlin, ``Secure state estimation and control for cyber security of the nonlinear power systems,'' {\it IEEE Trans. Control Netw. Syst.,} vol. 5, no. 3, pp. 1310--1321, 2018.

\bibitem{SW2012}
S. Wang, W. Gao, and A.S. Meliopoulos, ``An alternative method for power system dynamic
state estimation based on unscented transform,'' {\it IEEE Trans. power syst.}, vol. 27, no. 2, pp.942-950.

\bibitem{JK2019}
J. Kim, C. Lee, H. Shim, Y. Eun, and J. H. Seo, ``Detection of sensor attack and resilient state
estimation for uniformly observable nonlinear systems having redundant sensors,'' {\it IEEE Trans. Autom. Control,} vol. 64, no. 3, pp. 1162--1169, 2019.

\end{thebibliography}
\end{document}